\documentclass[journal]{IEEEtran}

\makeatletter
\def\ps@headings{%
	\def\@oddhead{\mbox{}\scriptsize\rightmark \hfil \thepage}%
	\def\@evenhead{\scriptsize\thepage \hfil \leftmark\mbox{}}%
	\def\@oddfoot{}%
	\def\@evenfoot{}}
\makeatother 

\IEEEoverridecommandlockouts

\usepackage{amsfonts}
\usepackage[dvips]{graphicx}
\usepackage{caption}
\usepackage{subfigure}
\usepackage{times}
\usepackage{cite}
\usepackage{lettrine}
\usepackage{amsmath}
\usepackage{amsmath}
\allowdisplaybreaks[4]
\usepackage{array}
\usepackage{amssymb}

\newcommand{\qed}{\hfill $\blacksquare$}

\usepackage{stfloats}
\usepackage{slashbox}
\usepackage{graphicx}
\usepackage{footnote}
\usepackage{booktabs}
\usepackage{array}
\usepackage{algorithmic}
\usepackage{algorithm}
\usepackage{subeqnarray}
\usepackage{cases}
\usepackage{threeparttable}
\usepackage{color}
\usepackage{bm}
\usepackage{bbm}

\DeclareMathOperator*{\argmin}{argmin}

\newtheorem{corollary}{\underline{Corollary}}[section]
\newtheorem{proposition}{Proposition}[section]

\begin{document}
	\bibliographystyle{IEEEtran}
	
	\title{Optimizing AI Service Placement and Resource Allocation in Mobile Edge Intelligence Systems}
	\IEEEoverridecommandlockouts
	
	\author{Zehong Lin,~\IEEEmembership{Student Member,~IEEE}, Suzhi Bi,~\IEEEmembership{Senior Member,~IEEE}, and Ying-Jun Angela Zhang,~\IEEEmembership{Fellow,~IEEE}
		
		\thanks{This article was presented in part at the IEEE Global Communications Conference (GLOBECOM), Taipei, Taiwan, December 2020 \cite{lin2020optimizing}.
			
			Z. Lin  and Y-J. A. Zhang are with the Department of Information Engineering, The Chinese University of Hong Kong, Hong Kong (e-mail: lz018@ie.cuhk.edu.hk; yjzhang@ie.cuhk.edu.hk).
			
			S. Bi is with the College of Electronics and Information Engineering, Shenzhen University, Shenzhen, China, 518060 (e-mail: bsz@szu.edu.cn). S. Bi is also with the Peng Cheng Laboratory, Shenzhen, China, 518066.}
	}
	
	\maketitle
	
	\begin{abstract}
		Leveraging recent advances on mobile edge computing (MEC), edge intelligence has emerged as a promising paradigm to support mobile artificial intelligence (AI) applications at the network edge. In this paper, we consider the AI service placement problem in a multi-user MEC system, where the access point (AP) places the most up-to-date AI program at user devices to enable local computing/task execution at the user side. To fully utilize the stringent wireless spectrum and edge computing resources, the AP sends the AI service program to a user only when enabling local computing at the user yields a better system performance. We formulate a mixed-integer non-linear programming (MINLP) problem to minimize the total computation time and energy consumption of all users by jointly optimizing the service placement (i.e., which users to receive the program) and resource allocation (on local CPU frequencies, uplink bandwidth, and edge CPU frequency). To tackle the MINLP problem, we derive analytical expressions to calculate the optimal resource allocation decisions with low complexity. This allows us to efficiently obtain the optimal service placement solution by search-based algorithms such as meta-heuristic or greedy search algorithms. To enhance the algorithm scalability in large-sized networks, we further propose an ADMM (alternating direction method of multipliers) based method to decompose the optimization problem into parallel tractable MINLP subproblems. The ADMM method eliminates the need of searching in a high-dimensional space for service placement decisions and thus has a low computational complexity that grows linearly with the number of users. Simulation results show that the proposed algorithms perform extremely close to the optimum and significantly outperform the other representative benchmark algorithms.
	\end{abstract}
	
	
	\section{Introduction}
	\IEEEPARstart{W}{ith} the rapid development of Internet of Things (IoT), tens of billions of mobile devices, like smartphones, wearable devices, and sensors, are connected to the Internet, generating unprecedented volumes of data, such as social media contents, mobile payment statistics, and users' geo-location information, at the network edge. This triggers the proliferation of various mobile artificial intelligence (AI) applications (e.g., augmented reality, autonomous driving, and intelligent personal assistants) to fully unleash the potential of mobile big data. Nonetheless, the intensive computational demand for training and inference of AI applications far exceeds the computation and energy capacity of mobile devices.
	
	Edge intelligence (EI) \cite{park_2019, EI_chen2019, wang_2020_2, shi_2020_2}, the integration of mobile edge computing (MEC) and AI technologies, has recently emerged as a promising paradigm to support computation-intensive AI applications at the network edge. Specifically, the edge servers of mobile networks, e.g., cellular base stations (BSs) and wireless access points (APs) \cite{MEC_survey, MEC_vision}, can provide cloud-like computing capabilities, greatly complementing the limited capacity of resource-constrained mobile devices. As the edge servers are in close proximity to the mobile devices and data sources, MEC avoids moving big data across the backhaul network compared with the conventional mobile cloud computing (MCC), and thus achieves lower latency and better privacy protection. With the aid of MEC, EI can push the computationally intensive training and inference processes of the AI models to the edge servers, making the mobile AI applications much more efficient. Meanwhile, recent advances in mobile AI chips, such as the neural processing units (NPU) integrated in HiSilicon's Kirin 970 chips and Apple's A11 bionic chips, equip the latest models of mobile devices with AI computation capabilities. With well-trained models, these advanced mobile devices can choose to run AI inference locally or at the edge servers \cite{EI_chen2019, wang_2020_2, shi_2020_2} following the two basic computation offloading models of MEC, i.e., binary offloading and partial offloading \cite{MEC_survey}.
	
	In recent years, joint optimization of computation offloading and system-level resource allocation (e.g., radio spectrum, computing power and transmit power) for MEC systems has attracted significant research interests \cite{MEC_zhang, MEC_wang, MEC_chen, MEC_you, MEC_bi, MEC_bi2, MEC_bi4, MEC_feng1, MEC_yan}. Most of the works implicitly assume that the required service programs for task computation are already available at both the edge servers and mobile devices. This assumption, however, is not true in AI services since the underlying AI models typically require continuous re-training. In particular, the AI models are trained using historical data and applied to the inference for future unseen data sampled from the same underlying distribution. Nonetheless, the environments are often nonstationary, and the data distribution can change over time. For instance, in an online shopping application, the customers' buying preferences may vary with time, depending on many factors, including the date, the availability of alternatives, etc. The changes of the underlying data distribution may result in concept drift problems \cite{drift1, drift2, drift3}, degrading AI inference performance. Therefore, to avoid model degradation over time, an AI service program must be updated either periodically or upon significant changes of the environment by re-training the AI model with newly collected data \cite{periodic_upload, D2D_sharing3}. The updated AI service program is then selectively disseminated to the edge servers and/or mobile devices. Notice that a server or a device can execute an AI task only when the updated service program is placed at it. Otherwise, its computation tasks must be offloaded to other devices where the service program is available. In this regard, \cite{infocom_2019} studied service placement in an MEC network with multiple edge servers to maximize the number of served computation offloading requests under edge storage, computation, and communication constraints. To cope with the unknown and fluctuating service demand, \cite{Xu_TWC} proposed an online learning algorithm to optimize spatial-temporal dynamic service placement decisions among multiple edge servers to minimize the computation delay. Considering parallel computing at both cloud and edge servers, \cite{Xu_infocom} and \cite{Xu_TMC} studied collaborative service placement and computation offloading to minimize the computation latency.
	
	The above works \cite{infocom_2019, Xu_TWC, Xu_infocom, Xu_TMC} assumed that mobile devices offload all their computation tasks for remote execution, and mainly focused on optimizing the service placement at the edge servers to ease the burden of the cloud. Nonetheless, due to the time-varying characteristic of wireless channels and limited edge computing capability, offloading all the computation tasks to the edge server is not always the optimal choice. To make efficient use of idle local computing power, it is more advantageous to opportunistically offload computation tasks and allow the mobile devices to execute some tasks locally \cite{MEC_zhang, MEC_wang, MEC_chen, MEC_you, MEC_bi, MEC_bi2, MEC_bi4, MEC_feng1, MEC_yan}. Noticeably, placing service programs at the mobile devices incurs additional program transmission delay. Recent work in \cite{MEC_bi3_2} takes such delay into account and jointly optimizes the service placement and computation offloading decisions in a single-user MEC system. The optimal design becomes much more complicated in a general multi-user scenario with heterogeneous wireless channel conditions and hardware configurations, where the users share limited system resources including the computing power of the server and the uplink channel spectrum for task offloading. In this case, the system resource allocation, computation offloading, and service placement decisions are closely correlated. For instance, whether placing the service at a user depends on the delay of obtaining the program, the user's local computing capability, the computing capability of the server, and the allocated bandwidth for task offloading. Therefore, we need to jointly optimize these coupling factors to achieve the optimal system computing performance.
	
	\begin{figure}[t]
		\centering
		\includegraphics[scale=0.6]{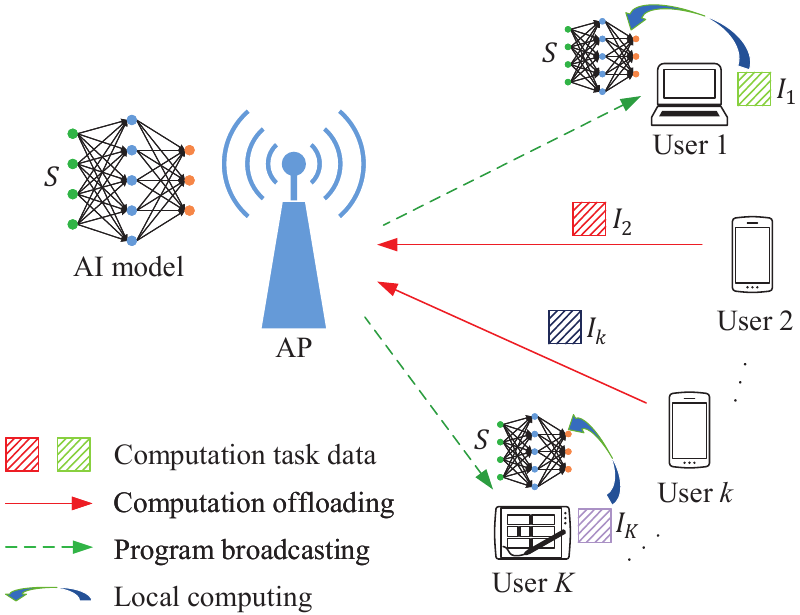}
		\caption{The considered MEC system with AI service placement.}  \vspace{-10pt}
		\label{fig:model}
	\end{figure}
	
	In this paper, we consider the AI service placement problem in a multi-user MEC system, as shown in Fig. \ref{fig:model}. Upon the update of an AI model, the edge server selectively transmits the program of the AI model to a subset of users via a broadcast channel. In particular, the AP sends the AI service program to a user only when enabling local computing at the user yields a better system performance in terms of the total time and energy consumption (TEC). We are interested in minimizing the total TEC of all users. The main contributions of this paper are summarized as follows.
	\begin{itemize}
		\item We formulate a mixed-integer non-linear programming (MINLP) problem for joint optimization of service placement, computational and radio resource allocation to minimize the total TEC of the users. The problem is challenging to solve due to the combinatorial service placement decision and its strong coupling with the communication and computational resource allocation decisions.
		
		\item We derive analytical expressions to efficiently calculate the optimal resource allocation decisions, including the local CPU frequencies, the edge CPU frequency, and the uplink bandwidth allocation, given the service placement decision. The analysis allows us to search for the optimal service placement solution at a low computational complexity via, e.g., greedy search or meta-heuristic methods.
		
		\item To avoid high-dimensional search when the network size is large, we propose an ADMM (alternating direction method of multipliers) based algorithm that decomposes the original problem into parallel and tractable subproblems, one for each user. As such, the total computational complexity of the ADMM-based algorithm increases linearly with the number of users, and is much more scalable than the search-based algorithms especially when the network size is large.
	\end{itemize}
	
	Simulation results show that the proposed algorithms achieve a close-to-optimal performance and significantly reduce the total computation delay and energy consumption compared with various benchmark algorithms. Moreover, we observe that the proposed search-based algorithms and the ADMM algorithm have their respective advantages. In particular, the search-based algorithms achieve lower computational complexity when the network size is small (e.g., $\leq 8$ users) due to the analytical expressions for calculating the optimal resource allocation decisions. On the other hand, the ADMM algorithm is preferred when the network size becomes large.
	
	The rest of the paper is organized as follows. We introduce the system model and formulate the joint optimization problem in Section II. In Section III, we derive analytical expressions to calculate the optimal computational and communication resource allocation decisions. Based on the analysis, we optimize the service placement decision via search-based algorithms. In Section IV, we propose an ADMM-based algorithm to enhance the algorithm scalability in large-sized networks. In Section V, we evaluate the proposed algorithms via extensive simulations. Finally, we conclude the paper in Section VI.
	
	\section{System Model and Problem Formulation}
	
	\subsection{System Model}
	As shown in Fig. \ref{fig:model}, we consider a multi-user MEC system consisting of $K$ single-antenna users, denoted by the set $\mathcal{K} = \{1, \cdots, K\}$, and a single-antenna AP co-located with an edge server.\footnote{In the remainder of this paper, we use AP and edge server interchangeably.} Suppose that each user has a certain amount of local data (e.g., personal images) to be processed by a common AI service (e.g., an image recognition program). The edge server periodically re-trains the AI model based on the latest data to avoid model degradation. It selectively disseminates the updated model to the mobile devices based on the optimal service placement decision. A mobile device can either offload its local data to the edge server for remote processing (e.g., AI inference) or process its data locally if it receives the service program from the edge server. One example application of on-device AI inference is skin cancer detection \cite{example1}, which deploys a pre-trained convolutional neural network (CNN) on a mobile device. For a new skin image provided by the mobile device, the AI model is used to classify skin lesions locally. Another application is smart classrooms in \cite{example2}, where three pre-trained deep neural network (DNN) models, for object, text, and voice recognition, are embedded in a mobile app to facilitate on-device DNN inference to control different classroom devices via an IoT micro-server.
	
	To reduce the communication overhead incurred by periodic service placement, we assume that the AP disseminates the service program via downlink broadcasting. Let $\mathcal{K}_1 \subseteq \mathcal{K}$ denote the subset of users the AP chooses to transmit the service program to, and $\mathcal{K}_0 = \mathcal{K} \setminus \mathcal{K}_1$ denote the set of remaining users. Notice that only the users in $\mathcal{K}_1$ are able to compute their tasks locally, while the users in $\mathcal{K}_0$ have to offload all the computation to the edge server for remote execution. We can show that the AP sends the program to a user only when the user is going to perform local computing. Otherwise, it may cause an unnecessary increase in service placement delay. Therefore, we suppose that all users in $\mathcal{K}_1$ perform local computing hereafter.
	
	In this paper, we consider a frequency division duplexing (FDD) operating mode where the downlink program broadcasting and the uplink computation o\textsc{}ffloading are operated simultaneously over orthogonal frequency bands, denoted by $W_{\text{D}}$ and $W_{\text{U}}$, respectively. Besides, the users in $\mathcal{K}_0$ share the uplink bandwidth using frequency division multiple access (FDMA). That is, user $k$ occupies a bandwidth of $a_k W_{\text{U}}$, where $a_k \in [0, 1]$ and $\sum_{k \in \mathcal{K}_0} a_k \leq 1$. Likewise, let $h_k$ and $g_k$ denote the wireless channel gains between the AP and user $k$ in the downlink and uplink, respectively. In this paper, we consider an offline model that the AP is assumed to have non-causal knowledge of users' channel state information (CSI) \footnote{ Similar to conventional wireless communication systems, the CSI can be obtained by channel estimation using pilot signals. If the AP suffers from CSI estimation errors, the performance of the proposed algorithms may degrade. In this case, some robust optimization techniques are needed to maintain the performance, which, however, is out of the scope of this paper.} and computation requirements.
	
	\subsection{Downlink Service Placement Model}
	Suppose that the AP disseminates a service program of size $S$ to the users in the set $\mathcal{K}_1$ by broadcasting in the downlink channel with power $p_0$. To ensure correct decoding of all users in $\mathcal{K}_1$, the AP adapts its broadcasting rate $r_0$ according to the worst-case user in $\mathcal{K}_1$. That is,
	\begin{align}
		r_0 = W_{\text{D}} \log_2 \bigg(1 + \frac{p_0 h_{\text{min}}}{W_{\text{D}} N_0} \bigg),  \label{downlink_rate}
	\end{align}
	where $h_{\text{min}} = \min\{h_k | k \in \mathcal{K}_1\}$ denotes the smallest downlink channel gain in the user set $\mathcal{K}_1$ and $N_0 $ denotes the noise power spectral density. Then, the time consumed on broadcasting the service program is $\tau_0 = \frac{S}{r_0}$.
	
	Let $p_k^r$ denote the circuit power consumption at the receiver of user $k$. The energy consumed for receiving the service program at user $k$ is
	\begin{align}
		e_k^r = p_k^r \tau_0.
	\end{align}	
	
	\subsection{Computation Model}	
	Suppose that user $k$ has $I_k$-bit task data to be computed by the AI service program.\footnote{The task data and service program are determined by the specific AI application. For the deep learning based image recognition application in \cite{image_recognition} that was written in C++, the task data can be personal images and the service program is the executable file compiled from the C++ code.} Besides, let $L_k$ denote the computing workload in terms of the total number of CPU cycles required for completing the task of user $k$. Depending on the local availability of the program, we describe the details of the local computing and edge computing models in the following.
	
	\subsubsection{Local Computing}
	A user $k$ conducts local computing when it is in $\mathcal{K}_1$, i.e., having the service program placed locally. Let $f_k^l$ denote the local CPU frequency of user $k$, which is limited by a maximum value $F_k$, i.e., $f_k^l \leq F_{k}$. Then, the time consumed on local computing by user $k$ is
	\begin{align}
		\tau_k^l = \frac{L_k}{f_k^l}.  \label{local_delay}
	\end{align}
	The corresponding energy consumption is \cite{MEC_wang}
	\begin{align}
		e_k^l = \kappa_k (f_k^l)^3 \tau_k^l = \kappa_k (f_k^l)^2 L_k, \label{local_energy}
	\end{align}
	where $\kappa_k > 0$ denotes the computing energy efficiency coefficient.
	
	\subsubsection{Edge Computing}
	A user $k$ offloads its computation tasks to the edge server when it is in $\mathcal{K}_0$. Let $p_{k}$ denote the transmit power of user $k$. Then, the uplink data rate of user $k$ is
	\begin{align}
		r_{k}^u = a_{k} W_{\text{U}} \log_2 \bigg(1 + \frac{p_{k} g_{k}}{a_{k} W_{\text{U}} N_0} \bigg). \label{upload_rate}
	\end{align}
	The time spent on offloading the task data of user $k$ is
	\begin{align}
		\tau_k^u = \frac{I_k}{r_{k}^u}, \label{edge_delay}
	\end{align}
	and the corresponding energy consumption is
	\begin{align}
		e_{k}^u = p_k \tau_k^u = p_k \frac{I_k}{r_{k}^u}.  \label{edge_energy}
	\end{align}
	
	Suppose that the edge server assigns a CPU frequency $f_k^c$ to compute the task of user $k$. Then, the task processing time of user $k$ at the edge server is
	\begin{align}
		\tau_k^c = \frac{L_k}{f_k^c},  ~~\forall k \in \mathcal{K}_0.
	\end{align}
	Due to the limitation of the computation capability of the edge server, the following edge CPU frequency constraint holds:
	\begin{align}
		\sum_{k \in \mathcal{K}_0} f_k^c \leq F^c,
	\end{align}
	where $F^c$ is the maximum CPU frequency of the edge server.
	
	In practice, the transmit power of the AP is much stronger than the users. Meanwhile, the size of the computation result is usually much smaller than the input data size. Thus, we can neglect the time spent on downloading the computing results from the AP to the users (as in \cite{MEC_zhang, MEC_you, MEC_feng1, MEC_bi, MEC_bi2}).
	
	\subsection{Problem Formulation}
	From the above discussion, we can calculate the total time consumption of user $k$ as
	\begin{align}
		T_k = \begin{cases}
			\tau_0 + \tau_k^l, & \mbox{if $k \in \mathcal{K}_1$},  \\
			\tau_k^u + \tau_k^c, & \mbox{if $k \in \mathcal{K}_0$},
		\end{cases} \label{delay_1}
	\end{align}
	and the total energy consumption of user $k$
	\begin{align}
		E_k = \begin{cases}
			e_k^r + e_k^l, & \mbox{if $k \in \mathcal{K}_1$}, \\
			e_{k}^u, & \mbox{if $k \in \mathcal{K}_0$}.
		\end{cases} \label{energy_1}
	\end{align}
	
	In particular, the total computation time of a user in $\mathcal{K}_1$ consists of the downlink service deployment delay and the local computing time. Likewise, the total computation time of a user in $\mathcal{K}_0$ consists of the task offloading time and the edge computing time. Define $\text{TEC}_k$ of user $k$ as the weighted sum of the computation time and energy consumption, i.e., $\text{TEC}_k = \beta_k^T T_k + \beta_k^E E_k$, where $\beta_k^T \geq 0$ and $\beta_k^E \geq 0$ are the weighting factors that satisfy $\beta_k^E = 1 - \beta_k^T$ \cite{MEC_bi3_2, TEC_ref}. We are interested in minimizing the total TEC, given by $\sum_{k \in \mathcal{K}} \text{TEC}_k$, by jointly optimizing the service placement decision $\mathcal{K}_1 \subseteq \mathcal{K}$, the local CPU frequencies $\mathbf{f}^l \triangleq \{f_k^l\}$, the uplink bandwidth allocation $\mathbf{a} \triangleq \{a_{k}\}$, and the edge CPU frequency allocation $\mathbf{f}^c \triangleq \{f_k^c\}$.
	Mathematically, the total TEC minimization problem is formulated as
	\begin{subequations}
		\begin{align}
			\mbox{(P1): } \min_{\mathcal{K}_1, \mathbf{f}^l, \mathbf{a}, \mathbf{f}^c} &V(\mathcal{K}_1, \mathbf{f}^l, \mathbf{a}, \mathbf{f}^c) \triangleq
			\sum_{k \in \mathcal{K}_1} \bigg[\beta_k^T \bigg(\tau_0 + \frac{L_k}{f_k^l} \bigg)  \nonumber \\
			&~+ \beta_k^E \left( p_k^r \tau_0 + \kappa_k (f_k^l)^2 L_k \right) \bigg]   \nonumber \\
			&~+ \sum_{k \in \mathcal{K}_0} \bigg[ \beta_k^T \bigg(\tau_k^u + \frac{L_k}{f_k^c} \bigg) + \beta_k^E p_k \tau_k^u \bigg] \label{obj1}  \\
			{\rm s.t.}~~~&\sum_{k  \in \mathcal{K}_0} a_{k} \leq 1, \label{cons1_1}  \\
			&\sum_{k \in \mathcal{K}_0} f_{k}^c \leq F^c, \label{cons1_2}  \\
			&0 \leq f_k^l \leq F_{k}, ~~\forall k \in \mathcal{K}_1, \label{cons1_3}   \\
			&a_{k} \geq 0, ~f_k^c \geq 0,  ~~\forall k \in \mathcal{K}_0, \label{cons1_4}  \\
			&\mathcal{K}_1 \subseteq \mathcal{K}, ~\mathcal{K}_0 = \mathcal{K} \setminus \mathcal{K}_1. \label{cons1_5}
		\end{align}
	\end{subequations}
	Define $V^*(\mathcal{K}_1) = \min\limits_{\mathbf{f}^l, \mathbf{a}, \mathbf{f}^c} V(\mathcal{K}_1, \mathbf{f}^l, \mathbf{a}, \mathbf{f}^c)$ as the optimal objective function value of (P1) given $\mathcal{K}_1$. Here, \eqref{cons1_1} and \eqref{cons1_2} correspond to the bandwidth allocation constraint and the edge CPU frequency allocation constraint, respectively. \eqref{cons1_3} is due to the local CPU frequency constraints.
	
	Problem (P1) is a mixed-integer non-linear programming (MINLP) problem, which is in general non-convex. In Section III and IV, we propose low-complexity algorithms to address the problem.

	\section{Optimal Computational and Communication Resource Allocation}
	We observe that (P1) is jointly convex in $(\mathbf{f}^l, \mathbf{a}, \mathbf{f}^c)$ once $\mathcal{K}_1$ is given. In this section, we derive analytical expressions to efficiently calculate the optimal computational and communication resource allocation $\{(\mathbf{f}^l)^*, \mathbf{a}^*, (\mathbf{f}^c)^*\}$ for a given service placement decision $\mathcal{K}_1$. Based on the analysis, search-based algorithms such as meta-heuristic methods (e.g., Gibbs sampling, particle swarm optimization, etc.) can be conducted to optimize $\mathcal{K}_1$ with low-complexity.
	
	\subsection{Optimal Resource Allocation for Given $\mathcal{K}_1$}
	Suppose that $\mathcal{K}_1$ is given. We can accordingly obtain $\mathcal{K}_0 = \mathcal{K} \setminus \mathcal{K}_1$. From \eqref{upload_rate} and \eqref{edge_delay}, we see that $\tau_k^u$ is uniquely determined by $a_k$. Therefore, it is equivalent to regard $\tau_k^u$'s as the optimization variables of (P1)
	and introduce the following constraints on $\tau_k^u$'s to (P1):
	\begin{align}
		\frac{I_k}{\tau_k^u} &\leq a_{k} W_{\text{U}} \log_2 \bigg(1 + \frac{p_{k} g_{k}}{a_{k} W_{\text{U}} N_0} \bigg), ~~\forall k \in \mathcal{K}_0, \label{cons1_6} \\
		\tau_k^u &\geq 0, ~~\forall k \in \mathcal{K}_0. \label{cons1_7}
	\end{align}
	
	Notice that (P1) can be separately optimized for the users in $\mathcal{K}_1$ and the users in $\mathcal{K}_0$. In particular, each user $k \in \mathcal{K}_1$ independently optimizes its local CPU frequency $f_k^l$ by solving
	\begin{align}
		\min_{0 \leq f_k^l \leq F_{k}} ~~& \beta_k^T \bigg(\tau_0 + \frac{L_k}{f_k^l} \bigg) + \beta_k^E \left( p_k^r \tau_0 + \kappa_k (f_k^l)^2 L_k \right).  \label{K1_opt}
	\end{align}
	Since \eqref{K1_opt} is a convex optimization problem, we can obtain the optimal solution $(f_k^l)^*$ by finding the stationary point and considering the boundary condition:
	\begin{align}
		(f_k^l)^* = \min \Bigg\{ F_k, \sqrt[3]{\frac{\beta_k^T}{2 \beta_k^E \kappa_k}} \Bigg\}.  \label{optimal_fkl}
	\end{align}
	From \eqref{optimal_fkl}, we observe that $(f_k^l)^*$ increases with the ratio $\frac{\beta_k^T}{\beta_k^E \kappa_k}$. Noticeably, a smaller ratio indicates more emphasis on minimizing the energy consumption at user $k$. Thus, the optimal local CPU frequency of user $k$ is reduced.
	
	On the other hand, for the users in $\mathcal{K}_0$, we need to jointly optimize the uplink time allocation $\{\tau_k^u\}$, the uplink bandwidth allocation $\mathbf{a}$, and the edge CPU frequency allocation $\mathbf{f}^c$ by solving
	\begin{subequations}
		\begin{align}
			\min_{\{\tau_k^u\}, \mathbf{a}, \mathbf{f}^c}~& \sum_{k \in \mathcal{K}_0} \bigg[ \beta_k^T \bigg(\tau_k^u + \frac{L_k}{f_k^c} \bigg) + \beta_k^E p_k \tau_k^u \bigg]  \label{obj2} \\
			{\rm s.t.}~~~~&\frac{I_k}{\tau_k^u} \leq a_{k} W_{\text{U}} \log_2 \bigg(1 + \frac{p_{k} g_{k}}{a_{k} W_{\text{U}} N_0} \bigg),  ~~\forall k \in \mathcal{K}_0,  \label{cons2_1}   \\
			&\sum_{k  \in \mathcal{K}_0} a_{k} \leq 1, \label{cons2_2}  \\
			&\sum_{k \in \mathcal{K}_0} f_{k}^c \leq F^c, \label{cons2_3}  \\
			&\tau_k^u \geq 0, ~a_{k} \geq 0, ~f_k^c \geq 0,  ~~\forall k \in \mathcal{K}_0. \label{cons23_4}
		\end{align}\label{problem2}
	\end{subequations}
	We can express the partial Lagrangian as
	\begin{align}
		\mathcal{L}&(\{\tau_k^u\}, \mathbf{a}, \mathbf{f}^c, \bm{\lambda}, \mu, \nu) = \sum_{k \in \mathcal{K}_0} \bigg[ \beta_k^T \bigg(\tau_k^u + \frac{L_k}{f_k^c} \bigg) + \beta_k^E p_k \tau_k^u \bigg]  \nonumber \\
		&\hspace{3.2em} + \sum_{k \in \mathcal{K}_0} \lambda_k \Bigg[ \frac{I_k}{\tau_k^u} - a_{k} W_{\text{U}} \log_2 \bigg(1 + \frac{p_{k} g_{k}}{a_{k} W_{\text{U}} N_0} \bigg) \Bigg]  \nonumber \\
		&\hspace{3.2em} + \mu \bigg(\sum_{k \in \mathcal{K}_0} a_{k} - 1 \bigg) + \nu \bigg(\sum_{k \in \mathcal{K}_0} f_k^c - F^c \bigg),  \label{Lagrangian_gamma}
	\end{align}
	where $\bm{\lambda} \triangleq \{\lambda_k\} \geq 0$ denotes the dual variables associated with the constraints in \eqref{cons2_1}. $\mu \geq 0$ and $\nu \geq 0$ are the dual variables associated with the constraints in \eqref{cons2_2} and \eqref{cons2_3}, respectively. Accordingly, the dual function is
	\begin{align}
		g(\bm{\lambda}, \mu, \nu) = \min_{\{\tau_k^u\}, \mathbf{a}, \mathbf{f}^c} ~&\mathcal{L}(\{\tau_k^u\}, \mathbf{a}, \mathbf{f}^c, \bm{\lambda}, \mu, \nu)  \nonumber \\
		{\rm s.t.} ~~~~&\tau_k^u \geq 0, ~a_{k} \geq 0, ~f_k^c \geq 0,  ~~\forall k \in \mathcal{K}_0, \label{dual_function2}
	\end{align}
	and the corresponding dual problem is
	\begin{align}
		\max_{\bm{\lambda}\geq 0, \mu \geq 0, \nu \geq 0}~& g(\bm{\lambda}, \mu, \nu).  \label{dual_problem2}
	\end{align}
	\eqref{problem2} is a convex problem, and thus strong duality holds between \eqref{problem2} and \eqref{dual_problem2}. Therefore, we can equivalently solve \eqref{problem2} by solving \eqref{dual_problem2}.
	
	Let $\{\lambda_k^*, \mu^*, \nu^* \}$ denote the optimal dual variables. Then, the closed-form expressions of the optimal solution $\{(\tau_k^u)^*, a_k^*, (f_k^c)^*\}$ to \eqref{problem2} are given in the following propositions.
	\begin{proposition}
		\label{proposition3_1}
		The optimal offloading time allocation $(\tau_k^u)^*$ is given by
		\begin{align}
			(\tau_k^u)^* = \sqrt{\frac{\lambda_k^* I_k}{\beta_k^T + \beta_k^E p_k}}, ~~\forall k \in \mathcal{K}_0.  \label{optimal_tau_k}
		\end{align}
	\end{proposition}
	\begin{IEEEproof}
		The partial derivative of $\mathcal{L}(\{\tau_k^u\}, \mathbf{a}, \mathbf{f}^c, \bm{\lambda}, \mu, \nu)$ with respect to $\tau_k^u$ is
		\begin{align}
			\frac{\partial \mathcal{L}(\{\tau_k^u\}, \mathbf{a}, \mathbf{f}^c, \bm{\lambda}, \mu, \nu)}{\partial \tau_k^u} = \beta_k^T + \beta_k^E p_k - \lambda_k \frac{I_k}{(\tau_k^u)^2}.
		\end{align}
		By setting $\frac{\partial \mathcal{L}(\{\tau_k^u\}, \mathbf{a}, \mathbf{f}^c, \bm{\lambda}, \mu, \nu)}{\partial \tau_k^u} = 0$ at the minimum point, we have
		\begin{align}
			\tau_k^u = \sqrt{\frac{\lambda_k I_k}{\beta_k^T + \beta_k^E p_k}},
		\end{align}
		which completes the proof.
	\end{IEEEproof}
	
	From Proposition \ref{proposition3_1}, we observe that a smaller value of $(\beta_k^T + \beta_k^E p_k)$ and/or a larger task data size $I_k$ leads to a longer offloading delay. Besides, $\lambda_k^* > 0$ must hold, because otherwise $(\tau_k^u)^* = 0$ and $r_k^u = \frac{I_k}{(\tau_k^u)^*} \rightarrow \infty$, which is not achievable.
	\begin{proposition} \label{proposition3_2}
		The optimal uplink bandwidth allocation $a_k^*$ is given by
		\begin{align}
			a_k^* = \frac{\frac{p_k g_k}{W_{\text{U}} N_0}}{- \bigg(W \bigg( - \frac{1}{\exp \big( \frac{\mu^* \ln2}{\lambda_k^* W_{\text{U}} } + 1 \big)} \bigg) \bigg)^{-1} - 1}, ~~\forall k \in \mathcal{K}_0,  \label{optimal_ak}
		\end{align}
		where $W(x)$ denotes the Lambert-W function, which is the inverse function of $z\exp(z) = x$, i.e., $z = W(x)$.
	\end{proposition}
	\begin{IEEEproof}
		The partial derivative of $\mathcal{L}(\{\tau_k^u\}, \mathbf{a}, \mathbf{f}^c, \bm{\lambda}, \mu, \nu)$ with respect to $a_k$ is
		\begin{align}
			&\frac{\partial \mathcal{L}(\{\tau_k^u\}, \mathbf{a}, \mathbf{f}^c, \bm{\lambda}, \mu, \nu)}{\partial a_k}  \nonumber \\
			=& - \frac{\lambda_k W_{\text{U}} }{\ln2} \bigg[ \ln \bigg(1 + \frac{p_k g_k}{a_k W_{\text{U}} N_0} \bigg) - \frac{p_k g_k}{a_k W_{\text{U}} N_0 + p_k g_k} \bigg] + \mu.
		\end{align}
		By setting $\frac{\partial \mathcal{L}(\{\tau_k^u\}, \mathbf{a}, \mathbf{f}^c, \bm{\lambda}, \mu, \nu)}{\partial a_k} = 0$ at the minimum point, we have
		\begin{align}\label{derivative_a}
			\ln \bigg(1 + \frac{p_k g_k}{a_k W_{\text{U}} N_0} \bigg) = \frac{\mu \ln2}{\lambda_k W_{\text{U}} } + 1 - \frac{1}{1 + \frac{p_k g_k}{a_k W_{\text{U}} N_0}}.
		\end{align}
		By taking a natural exponential operation at both sides, we have
		\begin{align}\label{derivative_ae}
			\bigg(1 + \frac{p_k g_k}{a_k W_{\text{U}} N_0}\bigg) \exp \bigg(\frac{1}{1 + \frac{p_k g_k}{a_k W_{\text{U}} N_0}} \bigg) = \exp \bigg( \frac{\mu \ln2}{ \lambda_k W_{\text{U}}} + 1 \bigg).
		\end{align}
		Consider two positive values $u$ and $v$ that satisfy $\frac{1}{u} \exp(u) = v$, it holds that
		\begin{align}
			-u \exp(-u) = - \frac{1}{v}.  \label{Lambert-W}
		\end{align}
		Therefore, we have $u = - W(-\frac{1}{v})$, where $W(x)$ is the Lambert-W function, which is the inverse function of $z\exp(z) = x$, i.e., $z = W(x)$. Comparing \eqref{derivative_ae} and \eqref{Lambert-W}, we can infer that
		\begin{align}
			&\frac{1}{1 + \frac{p_k g_k}{a_k W_{\text{U}} N_0}} = - W \bigg( - \frac{1}{\exp \big( \frac{\mu \ln2}{\lambda_k W_{\text{U}} } + 1 \big)} \bigg)  \nonumber \\
			\Rightarrow ~&a_k = \frac{\frac{p_k g_k}{W_{\text{U}} N_0}}{- \bigg(W \bigg( - \frac{1}{\exp \big( \frac{\mu \ln2}{\lambda_k W_{\text{U}} } + 1 \big)} \bigg) \bigg)^{-1} - 1},
		\end{align}
		which completes the proof.
	\end{IEEEproof}
	
	Proposition \ref{proposition3_2} indicates that $\mu^* > 0$ must hold. Otherwise, $a_k^* \rightarrow \infty$, which is evidently not true at the optimum because $a_k^* \leq 1$ must hold. Since $\mu^* > 0$, and $\lambda_k^* > 0$ by Proposition \ref{proposition3_1}, we have $- \frac{1}{\exp \big( \frac{\mu^* \ln2}{ \lambda_k^* W_{\text{U}}} + 1 \big)} \in (-1/e, 0)$. Moreover, we have $W(x) \in (-1, 0)$ when $x \in (- 1/e, 0)$. Thus, the right-hand side of \eqref{optimal_ak} is strictly positive, i.e., $a_k^* > 0$.
	
	\begin{proposition} \label{proposition3_3}
		The optimal edge CPU frequency allocation $(f_k^c)^*$ is given by
		\begin{align}
			(f_k^c)^* = \sqrt{\frac{\beta_k^T L_k}{\nu^*}}, ~~\forall k \in\mathcal{K}_0.  \label{optimal_fkc}
		\end{align}
	\end{proposition}
	\begin{IEEEproof}
		The partial derivative of $\mathcal{L}(\{\tau_k^u\}, \mathbf{a}, \mathbf{f}^c, \bm{\lambda}, \mu, \nu)$ with respect to $f_k^c$ is
		\begin{align}
			\frac{\partial \mathcal{L}(\{\tau_k^u\}, \mathbf{a}, \mathbf{f}^c, \bm{\lambda}, \mu, \nu)}{\partial f_k^c} = -\beta_k^T \frac{L_k}{(f_k^c)^2} + \nu.
		\end{align}
		By setting $\frac{\partial \mathcal{L}(\{\tau_k^u\}, \mathbf{a}, \mathbf{f}^c, \bm{\lambda}, \mu, \nu)}{\partial f_k^c} = 0$ at the minimum point, we have
		\begin{align}
			f_k^c = \sqrt{\frac{\beta_k^T L_k}{\nu}},
		\end{align}
		which completes the proof.
	\end{IEEEproof}
	
	Proposition \ref{proposition3_3} indicates that $\nu^* > 0$ must hold to ensure a finite $(f_k^c)^*$. Meanwhile, we observe that $(f_k^c)^*$ increases with the weighting factor $\beta_k^T$ and the computing workload $L_k$. This means that the edge server allocates more computing power to speed up the computation of the users that have heavier computing workload (larger $L_k$) or emphasize more on the computation delay (larger $\beta^T_k$).
	
	Based on Propositions \ref{proposition3_1}-\ref{proposition3_3}, we can apply the ellipsoid method \cite{ellipsoid} to obtain the optimal $\{\bm{\lambda}^*, \mu^*, \nu^*\}$. The basic idea of the ellipsoid method is to iteratively generate a sequence of ellipsoids with decreasing volumes from an initial ellipsoid $\mathcal{E}^{(0)}$ that contains $\{\bm{\lambda}^*, \mu^*, \nu^*\}$ \cite{ellipsoid}. Specifically, we can use any $\{\bm{\lambda}, \mu, \nu\} \geq 0$ as the center of $\mathcal{E}^{(0)}$ and set the volume to be sufficiently large to contain $\{\bm{\lambda}^*, \mu^*, \nu^*\}$. At each iteration $t$, we update the dual variables $\{\bm{\lambda}, \mu, \nu\}$ using the following subgradients:
	\begin{align}\hspace{-1em}
		\Delta \lambda_k &= \frac{I_k}{\tau_k^u} - a_{k} W_{\text{U}} \log_2 	\bigg(1 + \frac{p_{k} g_{k}}{a_{k} W_{\text{U}} N_0} \bigg),  ~~\forall k \in \mathcal{K}_0,  \label{subgradient_lambda} \\
		\Delta \mu &= \sum_{k \in \mathcal{K}_0} a_k - 1,  	\label{subgradient_mu} \\
		\Delta \nu &= \sum_{k \in \mathcal{K}_0} f_k^c - F^c,  		\label{subgradient_nu}
	\end{align}
	and generate a new ellipsoid $\mathcal{E}^{(t)}$ of reduced volume that contains the corresponding half-space of $\mathcal{E}^{(t - 1)}$ \cite{ellipsoid}. The update of $\{\bm{\lambda}, \mu, \nu\}$ repeats until the specified stopping criterion \cite{ellipsoid} is met. Since \eqref{problem2} is convex, the ellipsoid method guarantees to converge to the optimal solution.

	Algorithm 1 illustrates the pseudo code of the algorithm to solve (P1) given $\mathcal{K}_1$. In this algorithm, the complexity of obtaining the optimal local CPU frequencies $(\mathbf{f}^l)^*$ is $O(|\mathcal{K}_1|)$, because we can directly calculate the optimal $(f_k^l)^*$ in closed-form for the users in $\mathcal{K}_1$. In addition, the ellipsoid method requires $O(m^2)$ iterations to converge, where $m$ is the number of dual variables \cite{cvx}. In this paper, $m = |\mathcal{K}_0| + 2$. Since we can calculate the optimal primal variables $\{(\tau_k^u)^*, a_k^*, (f_k^c)^*\}$ in closed-form, the complexity of each iteration in the ellipsoid method is proportional to the number of users in $\mathcal{K}_0$, i.e., $O(|\mathcal{K}_0|)$. Therefore, the overall computational complexity is $O\big( m^2 |\mathcal{K}_0| + |\mathcal{K}_1| \big)$. Since $|\mathcal{K}_0| \leq K$ and $|\mathcal{K}_1| = K - |\mathcal{K}_0|$, we can conclude that the overall computational complexity of Algorithm 1 is upper bounded by $O\big( K ^3 \big)$.
	\begin{algorithm}[t]
		\caption{Optimal Resource Allocation for Problem (P1) With Given $\mathcal{K}_1$}
		\begin{algorithmic}[1]
			\STATE \textbf{input:} service placement decision $\mathcal{K}_1$.
			\STATE \textbf{initialization:} $\mathcal{K}_0 = \mathcal{K} \setminus \mathcal{K}_1$, $\{\bm{\lambda}, \mu, \nu\} \geq 0$ for $\mathcal{K}_0$;
			\FOR{each $k \in \mathcal{K}_1$}
			\STATE Calculate $(f_k^l)^*$ using \eqref{optimal_fkl}.
			\ENDFOR
			\REPEAT
			\FOR{each $k \in \mathcal{K}_0$}
			\STATE Calculate $a_k^*$ using \eqref{optimal_ak}, and $(f_k^c)^*$ using \eqref{optimal_fkc};
			\ENDFOR
			\STATE Update $\{\bm{\lambda}, \mu, \nu\}$ by the ellipsoid method using the subgradients defined in \eqref{subgradient_lambda}-\eqref{subgradient_nu}.
			\UNTIL{$\{\bm{\lambda}, \mu, \nu\} \geq 0$ converge to a prescribed accuracy.}
			\RETURN the optimal $\{(\mathbf{f}^l)^*, \mathbf{a}^*, (\mathbf{f}^c)^*\}$ to (P1) given $\mathcal{K}_1$.
		\end{algorithmic}
	\end{algorithm}  \label{algorithm1}
	
	\subsection{Optimization of Service Placement Decision}
	The analysis in the previous subsection allows us to efficiently obtain the optimal resource allocation solutions $\{(\mathbf{f}^l)^*, \mathbf{a}^*, (\mathbf{f}^c)^*\}$ with given $\mathcal{K}_1$. This facilitates low-complexity implementation of search-based algorithms, such as the meta-heuristic algorithms including Gibbs sampling \cite{GS}, particle swarm optimization \cite{PS}, etc. to obtain the optimal $\mathcal{K}_1$. In particular, meta-heuristic algorithms strategically sample a subset of all feasible solutions of $\mathcal{K}_1$. With the closed-form expressions derived in Propositions \ref{proposition3_1}-\ref{proposition3_3}, we can quickly calculate the objective function value associated with each sampled $\mathcal{K}_1$ by Algorithm 1, thus significantly expediting the meta-heuristic algorithms.
	
	To further reduce the complexity, we discuss below an iterative greedy search algorithm. Let $\mathcal{K}_1^{(n)}$ denote the service placement decision at iteration $n$. Likewise, we have $\mathcal{K}_0^{(n)} = \mathcal{K} \setminus \mathcal{K}_1^{(n)}$. Correspondingly, the optimal objective value of (P1) at iteration $n$ is $V^*\big(\mathcal{K}_1^{(n)}\big)$. We initially set $\mathcal{K}_0^{(0)}= \mathcal{K}$ and $\mathcal{K}_1^{(0)} = \emptyset$. Then, in each iteration $n \geq 1$, we find the best user in $\mathcal{K}_{0}^{(n - 1)}$ such that once the user is removed from $\mathcal{K}_{0}^{(n - 1)}$ and assigned to $\mathcal{K}_{1}^{(n - 1)}$, the optimal total TEC $V^*\big(\mathcal{K}_1^{(n - 1)}\big)$ drops the most significantly. The process repeats until we cannot further decrease the total TEC by moving a user from $\mathcal{K}_0^{(n - 1)}$ to $\mathcal{K}_1^{(n - 1)}$, or $\mathcal{K}_0^{(n - 1)} = \emptyset$.
	
	There are at most $K$ iterations in the greedy search algorithm. In the $n$-th iteration, the algorithm needs to search over $(K - n + 1)$ users in the set $\mathcal{K}_0^{(n - 1)}$ and solves the corresponding optimization problem in (P1). Thus, a total of $\sum_{n = 1}^K K - n + 1 = \frac{K^2 + K}{2}$ optimization problems need to be solved in the worst case. As discussed, the complexity of Algorithm 1 is upper bounded by $O\big( K^3 \big)$. Thus, the overall complexity of the greedy search algorithm is upper bounded by $O\big(K^5 \big)$, implying that the algorithm can find the solution in polynomial time.
	
	\subsection{A Homogeneous Special Case}
	In this subsection, we study a special case where the users differ only by their wireless channel gains $g_k$'s and $h_k$'s. In this case, the weighting factors, local computing capability, computing energy efficiency, transmit power, circuit power consumption, and task parameters $(I_k, L_k)$ are identical for all users, i.e., $\beta_k^T = \beta$, $\beta_k^E = 1 - \beta$, $F_k = F$, $\kappa_k = \kappa$, $p_k = p$, $p_k^r = p^r$, $I_k = I$, and $L_k = L, \forall k \in \mathcal{K}$. In this case, $(f_k^l)^*$'s in \eqref{optimal_fkl} are equal at the optimum, i.e., $(f_k^l)^* = \min \left\{ F, \sqrt[3]{\frac{\beta}{2 (1 - \beta) \kappa}} \right\}$, for the users in $\mathcal{K}_1$. Consequently, the local computing delay and energy consumption are equal for all users in $\mathcal{K}_1$. Likewise, $(f_k^c)^*$'s are equal at the optimum for all users in $\mathcal{K}_0$. Since $\sum_{k \in \mathcal{K}_0} f_k^c = F^c$ must hold at the optimum, the total edge CPU frequency is equally allocated to the users in $\mathcal{K}_0$.
	
	On the other hand, we have the following proposition on the relation between $g_k$, $(\tau_k^u)^*$, the uplink spectral efficiency $\log_2 \big(1 + \frac{p g_k}{a_k^* W_{\text{U}} N_0} \big)$ and the offloading data rate $a_k^* W_{\text{U}} \log_2 \big(1 + \frac{p g_k}{a_k^* W_{\text{U}} N_0} \big)$.
	\begin{proposition}\label{proposition3_4}
		For a user in $\mathcal{K}_0$, a worse uplink channel gain $g_k$ results in a longer offloading time $(\tau_k^u)^*$, a lower spectral efficiency in the uplink, i.e., $\log_2 \big(1 + \frac{p g_k}{a_k^* W_{\text{U}} N_0} \big)$, and a smaller offloading data rate $a_k^* W_{\text{U}} \log_2 \big(1 + \frac{p g_k}{a_k^* W_{\text{U}} N_0} \big)$.
	\end{proposition}
	
	Before proving Proposition \ref{proposition3_4}, we first prove the following Corollary on the relation between the optimal $\lambda_k^*$ and $g_k$.
	\begin{corollary}\label{corollary3_1}
		The optimal $\lambda_k^*$ is a non-increasing function of $g_k$.
	\end{corollary}
	\begin{IEEEproof}
		We prove Corollary \ref{corollary3_1} by contradiction. Suppose that $\lambda_k^*$ increases with $g_k$. According to the KKT condition $\lambda_k^* \Big[\frac{I}{(\tau_k^u)^*} - a_k^* W_{\text{U}} \log_2 \big(1 + \frac{p g_{k}}{a_{k}^* W_{\text{U}} N_0} \big) \Big] = 0$ and $\lambda_k^* > 0$ from Proposition \ref{proposition3_1}, we have $\frac{I}{(\tau_k^u)^*} - a_k^* W_{\text{U}} \log_2 \big(1 + \frac{p g_{k}}{a_{k}^* W_{\text{U}} N_0} \big) = 0$ at the optimum for all $k \in \mathcal{K}_0$. Note that $(\tau_k^u)^*$ increases with $\lambda_k^*$ according to \eqref{optimal_tau_k}, and thus it also increases with $g_k$. In addition, since $W(x)$ is an increasing function when $x \in (-1/e, 0)$, we can infer from \eqref{optimal_ak} that $a_k^*$ increases with both $g_k$ and $\lambda_k^*$. Therefore, $\frac{I}{(\tau_k^u)^*} - a_k^* W_{\text{U}} \log_2 \big(1 + \frac{p g_{k}}{a_{k}^* W_{\text{U}} N_0} \big)$ decreases with $g_k$. Thus, the condition $\frac{I}{(\tau_k^u)^*} - a_k^* W_{\text{U}} \log_2 \big(1 + \frac{p g_{k}}{a_{k}^* W_{\text{U}} N_0} \big) = 0$ cannot be simultaneously satisfied for all $k \in \mathcal{K}_0$. This contradiction implies that the assumption must be false, which leads to the proof.
	\end{IEEEproof}
	
	\emph{Proof of Proposition \ref{proposition3_4}:} \eqref{optimal_tau_k} and Corollary \ref{corollary3_1} indicate that a user with a worse uplink channel condition consumes a longer time $(\tau_k^u)^*$ to offload its computation task at the optimum. Since the task data size is identical for all users in the homogeneous case, the offloading data rate $a_k^* W_{\text{U}} \log_2 \big(1 + \frac{p g_k}{a_k^* W_{\text{U}} N_0} \big)$ decreases when $g_k$ becomes worse according to \eqref{upload_rate} and \eqref{edge_delay}.
	
	In the following, we prove that the optimal uplink spectral efficiency also drops when $g_k$ is poor. Indeed, by substituting \eqref{optimal_ak} into \eqref{upload_rate}, we can express the optimal uplink spectral efficiency of user $k$ as
	\begin{align}
		\log_2 \bigg(1 + \frac{p g_k}{a_k^* W_{\text{U}} N_0} \bigg) = \log_2 \big( \chi_k(\lambda_k^*, \mu^*)\big),  \label{uplink_rate}
	\end{align}
	where
	\begin{align}
		\chi_k(\lambda_k^*, \mu^*) = - \bigg(W \bigg( - \frac{1}{\exp \big( \frac{\mu^* \ln2}{\lambda_k^* W_{\text{U}} } + 1 \big)} \bigg) \bigg)^{-1}
	\end{align}
	is a decreasing function in $\lambda_k^*$. We infer from \eqref{uplink_rate} that the spectral efficiency of the uplink from user $k$ to the AP decreases with $\lambda_k^*$, and thus increases with $g_k$. This completes the proof. \qed
	
	The above proposition shows that not only the uplink task offloading rate, but also the uplink spectral efficiency drops when $g_k$ becomes worse. Interestingly, we observe in Fig. \ref{fig:opt_p}(c) in the simulation section that the optimal uplink bandwidth allocation $a_k^*$ also increases when $g_k$ is smaller. This interesting phenomenon and Proposition \ref{proposition3_4} imply that the users with worse uplink channels will be allocated with more uplink bandwidth but still result in a lower uplink data rate. In other words, it is spectrally inefficient to let users with poor channel gains offload their tasks for edge computing.
	
	Inspired by the above analysis and observation, we design the following uplink-based heuristic algorithm for the homogeneous special case. In particular, we sort all users according to the ascending order of $g_k$, and initialize $\mathcal{K}_1 = \emptyset$. At iteration $l$, we select user $k_l$ with the $l$-th smallest channel gain. The user $k_l$ is added into $\mathcal{K}_1$, i.e., $\mathcal{K}_1 = \mathcal{K}_1 \cup \{k_l\}$, if it reduces the objective value $V^*(\mathcal{K}_1)$. The process repeats until $l = K$. This algorithm solves (P1) $K$ times. The complexity is significantly lower than the greedy search algorithm proposed in Section III-B, which needs to solve (P1) $O(K^2)$ times. We will later show in Section V-A that this heuristic algorithm performs extremely close to the optimal one in the homogeneous special case.
	
	\section{Joint Optimization Using ADMM-Based Algorithm}
	The complexity of the aforementioned search-based algorithms becomes high when $K$ grows large. In this section, we propose an ADMM-based algorithm to decompose (P1) into $K$ parallel MINLP problems, one for each user. As such, the overall complexity grows much more slowly when $K$ increases.
	
	We introduce binary decision variables $b_k$'s to denote the service placement decisions, where $b_k = 1$ if user $k \in \mathcal{K}_1$ and $b_k = 0$ if user $k \in \mathcal{K}_0$. Denote $\mathbf{b} \triangleq \{b_k\}$. In (P1) $\tau_0$ introduces strong coupling among the users, as it is determined by the worst channel gain among the users in $\mathcal{K}_1$. To facilitate the decomposition, we regard $\tau_0$ as an optimization variable and reformulate (P1) as
	\begin{subequations}
		\begin{align}
			\mbox{(P2): } \min_{\mathbf{b}, \mathbf{f}^l, \mathbf{a}, \mathbf{f}^c, \tau_0} &\sum_{k = 1}^K \Bigg\{ b_k \bigg[\beta_k^T \bigg(\tau_0 + \frac{L_k}{f_k^l} \bigg)   \nonumber \\
			&+ \beta_k^E \Big( p_k^r \tau_0  + \kappa_k (f_k^l)^2 L_k \Big) \bigg]  \nonumber \\
			&+ (1 - b_k) \bigg[ \beta_k^T \bigg(\frac{I_k}{r_{k}^u} + \frac{L_k}{f_k^c} \bigg) + \beta_k^E p_k \frac{I_k}{r_{k}^u} \bigg] \Bigg\} \label{obj3} \\
			{\rm s.t.}~~~~&\sum_{k = 1}^K a_{k} \leq 1, \label{cons3_1}  \\
			&\sum_{k = 1}^K f_{k}^c \leq F^c, \label{cons3_2}  \\
			&0 \leq f_k^l \leq F_{k}, ~~\forall k \in \mathcal{K}, \label{cons3_3}   \\
			&a_{k} \geq 0, ~f_k^c \geq 0,  ~~\forall k \in \mathcal{K}, \label{cons3_4} \\
			&b_k \in \{0, 1\}, ~~\forall k \in \mathcal{K},  \label{cons3_5} \\
			&\tau_0 \geq b_k \frac{S}{W_{\text{D}} \log_2 \Big(1 + \frac{p_0 h_{k}}{W_{\text{D}} N_0 } \Big)}, ~~\forall k \in \mathcal{K} \label{cons3_6}.
		\end{align}  \label{problem3}
	\end{subequations}
	
	Note that the optimization variables $\mathbf{a}$, $\mathbf{f}^c$ and $\tau_0$ are coupled among the users in the constraints \eqref{cons3_1}, \eqref{cons3_2} and \eqref{cons3_6}, respectively. To decompose (P2), we introduce the local copies of the variables $\mathbf{a}$, $\mathbf{f}^c$ and $\tau_0$ as $\mathbf{x} \triangleq \{x_k\}$, $\mathbf{y} \triangleq \{y_k\}$ and $\mathbf{z} \triangleq \{z_k\}$, respectively. Then, we reformulate (P2) as
	\begin{subequations}
		\begin{align}\hspace{-0.5em}
			\min_{ \mathbf{b}, \mathbf{f}^l, \mathbf{a}, \mathbf{f}^c, \tau_0, \mathbf{x}, \mathbf{y}, \mathbf{z} }\ & \sum_{k = 1}^K q_k(b_k, f_k^l, x_k, y_k, z_k) + g(\mathbf{a}, \mathbf{f}^c, \tau_0)  \\
			{\rm s.t.}~~~~~~~&\eqref{cons3_3}, \eqref{cons3_5},  \nonumber \\
			&z_k \geq b_k \frac{S}{W_{\text{D}} \log_2 \big(1 + \frac{p_0 h_{k}}{W_{\text{D}} N_0 } \big)}, ~~\forall k \in \mathcal{K},  \label{cons5_3}  \\
			&x_k = a_k, ~~\forall k \in \mathcal{K},  \label{cons5_4}  \\
			&y_k = f_k^c, ~~\forall k \in \mathcal{K},  \label{cons5_5} \\
			&z_k = \tau_0, ~~\forall k \in \mathcal{K},  \label{cons5_6} \\
			&x_k \geq 0, ~y_k \geq 0,  ~~\forall k \in \mathcal{K},  \label{cons5_7}
		\end{align} \label{problem5}
	\end{subequations}
	where
	\begin{align}
		q_k&(b_k, f_k^l, x_k, y_k, z_k) \nonumber \\
		~~&= b_k \bigg[ \beta_k^T \bigg(z_k + \frac{L_k}{f_k^l} \bigg) + \beta_k^E \left( p_k^r z_k + \kappa_k (f_k^l)^2 L_k \right) \bigg] \nonumber \\
		~~&~~~~+ (1 - b_k) \bigg[\beta_k^T \bigg(\frac{I_k}{{r_{k}^{u}}'} + \frac{L_k}{y_k} \bigg) + \beta_k^E \frac{p_k I_k}{{r_{k}^{u}}'} \bigg],
	\end{align}
	and ${r_{k}^{u}}' = x_k W_{\text{U}} \log_2 \Big(1 + \frac{p_k g_k}{x_k W_{\text{U}} N_0} \Big)$.
	Besides,
	\begin{align}
		g(\mathbf{a}, \mathbf{f}^c, \tau_0) =
		\begin{cases}
			0, ~~~~~~\mbox{if $(\mathbf{a}, \mathbf{f}^c, \tau_0) \in \mathcal{G}$}, \\
			+\infty, ~~\mbox{otherwise},
		\end{cases}  \label{g_v}
	\end{align}
	where
	\begin{align}
		\mathcal{G} = \Bigg\{(\mathbf{a}, \mathbf{f}^c, \tau_0) \bigg| \sum_{k = 1}^K a_{k} \leq 1, \sum_{k = 1}^K f_{k}^c &\leq F^c, a_{k} \geq 0, f_k^c \geq 0,  \nonumber  \\
		&k \in \mathcal{K}; \tau_0 \geq 0 \Bigg\}.  \nonumber
	\end{align}
	
	Now, we can apply the ADMM technique \cite{ADMM} to decompose Problem \eqref{problem5}. By introducing multipliers to the constraints in \eqref{cons5_4}-\eqref{cons5_6}, the augmented Lagrangian of \eqref{problem5} is
	\begin{align}
		&\mathcal{L}(\mathbf{u}, \mathbf{v}, \bm{\theta}) = \sum_{k = 1}^K q_k(\mathbf{u}) + g(\mathbf{v}) + \sum_{k = 1}^K \rho_k (x_k - a_k)  \nonumber \\
		&~~~+ \sum_{k = 1}^K \phi_k (y_k - f_k^c) + \sum_{k = 1}^K \varphi_k (z_k - \tau_0)  + \frac{c}{2} \sum_{k = 1}^K (x_k - a_k)^2  \nonumber \\
		&~~~+ \frac{c}{2} \sum_{k = 1}^K (y_k - f_k^c)^2 + \frac{c}{2} \sum_{k = 1}^K (z_k - \tau_0)^2,
	\end{align}
	where $\mathbf{u} = \{\mathbf{b}, \mathbf{f}^l, \mathbf{x}, \mathbf{y}, \mathbf{z}\}$, $\mathbf{v} = \{\mathbf{a}, \mathbf{f}^c, \tau_0\}$, $\bm{\theta} = \{\bm{\rho}, \bm{\phi}, \bm{\varphi}\}$, and $c > 0$ is a fixed step size. Accordingly, the dual function is
	\begin{align}
		d(\bm{\theta}) = \min_{\mathbf{u}, \mathbf{v}} ~&\mathcal{L}(\mathbf{u}, \mathbf{v}, \bm{\theta}) \nonumber \\
		{\rm s.t.} ~~&\eqref{cons3_3}, \eqref{cons3_5}, \eqref{cons5_3}, \eqref{cons5_7},
	\end{align}
	and the dual problem is
	\begin{equation}
		\max_{\bm{\theta}}~ d(\bm{\theta}).  \label{ADMM_dual}
	\end{equation}
	
	The ADMM method solves the dual problem \eqref{ADMM_dual} by iteratively updating $\mathbf{u}$, $\mathbf{v}$, and $\bm{\theta}$. We denote the values in the $i$-th iteration as $\{\mathbf{u}^i, \mathbf{v}^i, \bm{\theta}^i\}$. Then, in the $(i + 1)$-th iteration, the variables are updated sequentially as follows:
	\subsubsection{Step 1}
	In this step, we update the local variables $\mathbf{u}$ as
	\begin{align}
		\mathbf{u}^{i + 1} = \argmin_{\mathbf{u}} \mathcal{L} (\mathbf{u}, \mathbf{v}^i, \bm{\theta}^i).  \label{update_u}
	\end{align}
	Notice that the minimization problem in \eqref{update_u} can be decomposed into $K$ parallel subproblems. Each subproblem solves two optimization problems, one for $b_k = 0$ and another one for $b_k = 1$. In particular, the optimization problem for $b_k = 0$ is
	\begin{align}{{\hspace{-3pt}}}
		&\min_{x_k, y_k, z_k \geq 0} ~\beta_k^T \bigg( \frac{I_k}{{r_{k}^{u}}'} + \frac{L_k}{y_k} \bigg) + \beta_k^E \frac{p_k I_k}{{r_{k}^{u}}'} + \rho_k^i x_{k} + \phi_k^i y_k  \nonumber \\
		&~~~+ \varphi_k^i z_k + \frac{c}{2} \big( x_k - a_k^i \big)^2 + \frac{c}{2} (y_k - (f_k^c)^i)^2 + \frac{c}{2} (z_k - \tau_0^i)^2,
		\label{step1_0}
	\end{align}
	and the optimization problem for $b_k = 1$ is
	\begin{align}
		\min_{f_k^l, x_k, y_k, z_k} ~& \beta_k^T \bigg(z_k + \frac{L_k}{f_k^l} \bigg) + \beta_k^E \left( p_k^r z_k + \kappa_k (f_k^l)^2 L_k \right)  \nonumber \\
		& + \rho_k^i x_{k} + \phi_k^i y_k + \varphi_k^i z_k + \frac{c}{2} \big( x_k - a_k^i \big)^2  \nonumber \\
		& + \frac{c}{2} (y_k - (f_k^c)^i)^2 + \frac{c}{2} (z_k - \tau_0^i)^2  \nonumber \\
		{\rm s.t.} ~~~~~&0 \leq f_k^l \leq F_k,  \nonumber \\
		&z_k \geq \frac{S}{W_{\text{D}} \log_2 \big(1 + \frac{p_0 h_{k}}{W_{\text{D}} N_0 } \big)},  \nonumber \\
		&x_k \geq 0, ~y_k \geq 0. \label{step1_1}
	\end{align}
	
	Note that both \eqref{step1_0} and \eqref{step1_1} are strictly convex problems that can be solved using general convex optimization algorithm, e.g., projected Newton's method \cite{cvx}. Therefore, we can simply select $b_k = 0$ or $1$ that yields a smaller objective value as $b_k^{i + 1}$, and the corresponding optimal solution as $\{(f_k^l)^{i + 1}, x_k^{i + 1}, y_k^{i + 1}, z_k^{i + 1}  \}$. After solving the $K$ subproblems, the optimal solution to \eqref{update_u} is given by $\mathbf{u}^{i + 1} = \{\mathbf{b}^{i + 1}, (\mathbf{f}^l)^{i + 1}, \mathbf{x}^{i + 1}, \mathbf{y}^{i + 1}, \mathbf{z}^{i + 1} \}$. Therefore, the overall computational complexity of Step 1 is $O(K)$. Notice that the $K$ subproblems can be solved in parallel, thus the computational time of Step 1 is constant when we conduct parallel computing.
	
	\subsubsection{Step 2}
	Having obtained $\mathbf{u}^{i + 1}$, we update the global variables $\mathbf{v}$ as
	\begin{align}
		\mathbf{v}^{i + 1} &= \argmin_{\mathbf{v}} \mathcal{L} (\mathbf{u}^{i + 1}, \mathbf{v}, \bm{\theta}^i).  \label{update_v}
	\end{align}
	By the definition of $g(\mathbf{v})$ in \eqref{g_v}, $\mathbf{v}^{i + 1} \in \mathcal{G}$ must hold at the optimum. Accordingly, the minimization problem in \eqref{update_v} is equivalent to the following convex optimization problem
	\begin{align}
		\mathbf{v}^{i + 1} = \argmin_{\mathbf{a}, \mathbf{f}^c, \tau_0} &\sum_{k = 1}^K \rho_k^i (x_k^{i + 1} - a_k) + \sum_{k = 1}^K \phi_k^i (y_k^{i + 1} - f_k^c)  \nonumber \\
		& + \sum_{k = 1}^K \varphi_k^i (z_k^{i + 1} - \tau_0) + \frac{c}{2} \sum_{k = 1}^K (x_k^{i + 1} - a_k)^2  \nonumber \\
		& + \frac{c}{2} \sum_{k = 1}^K (y_k^{i + 1} - f_k^c)^2 + \frac{c}{2} \sum_{k = 1}^K (z_k^{i + 1} - \tau_0)^2  \nonumber \\
		{\rm s.t.} ~~&\sum_{k = 1}^K a_{k} \leq 1,  \nonumber \\
		&\sum_{k = 1}^K f_{k}^c \leq F^c,  \nonumber \\
		&a_{k} \geq 0, ~f_{k}^c \geq 0,  ~~\forall k \in \mathcal{K},  \nonumber \\
		&\tau_0 \geq 0.
		\label{step2}
	\end{align}
	Instead of applying standard convex optimization tools to solve \eqref{step2}, we propose a low-complexity algorithm in the following.
	
	Let $\psi$ and $\gamma$ denote the Lagrangian multipliers associated with constraints $\sum_{k = 1}^K a_{k} \leq 1$ and $\sum_{k = 1}^K f_{k}^c \leq F^c$, respectively. Then, we can obtain the optimal $\{\mathbf{a}^*, (\mathbf{f}^c)^*, \tau_0^*\}$ in closed-form as
	\begin{align}
		a_k^* &= \bigg( x_k^{i + 1} + \frac{\rho_k^i - \psi^*}{c} \bigg)^+,  ~~\forall k \in \mathcal{K},  \label{optimal_admm_a}\\
		(f_k^c)^* &= \bigg( y_k^{i + 1} + \frac{\phi_k^i - \gamma^*}{c} \bigg)^+,  ~~\forall k \in \mathcal{K}, \label{optimal_admm_fkc}
	\end{align}
	and
	\begin{align}\label{optimal_tau_0}
		\tau_0^* = \bigg(\frac{\sum_{k = 1}^{K} z_k^{i + 1}}{K} + \frac{\sum_{k = 1}^{K} \varphi_k^{i}}{c K} \bigg)^+,
	\end{align}
	where $(\cdot)^+ = \max\{\cdot, 0\}$. As $a_k^*$ is non-increasing with $\psi^* \geq 0$, we can obtain the optimal $\psi^*$ by bisection search over $\psi^* \in (0, \bar{\psi})$, where $\bar{\psi}$ is a sufficiently large value, until $\sum_{k = 1}^K a_{k}^* = 1$ is satisfied. Likewise, $(f_k^c)^*$ is non-increasing with $\gamma^* \geq 0$. Therefore, we can obtain the optimal $\gamma^*$ by bisection search over $\gamma^* \in (0, \bar{\gamma})$, where $\bar{\gamma}$ is a sufficiently large value. The pseudo-code of the algorithm is shown in Algorithm 2. Given an error tolerance $\epsilon_1$ for $\psi^*$, the bisection search for $\psi^*$ terminates within $\log_2(\frac{\bar{\psi}}{\epsilon_1})$ iterations. Likewise, the bisection search for $\gamma^*$ terminates within $\log_2(\frac{\bar{\gamma}}{\epsilon_2})$ iterations, where $\epsilon_2$ is the error tolerance for $\gamma^*$. Overall, the computational complexity of Step 2 is $O(K)$. In addition, we can calculate $a_k^*$'s and $(f_k^c)^*$'s for the $K$ users in parallel under given dual variables $\psi$ and $\gamma$, thus the computational time of Step 2 is constant when we conduct parallel computing.
	\begin{algorithm}[t]
		\caption{Bisection Search Algorithm for Solving Problem \eqref{step2}}
		\begin{algorithmic}[1]
			\STATE \textbf{input:} local variables $\mathbf{u}^{i + 1}$.
			\STATE \textbf{initialization:} $\varepsilon_1 = 10^{-4}$; $\varepsilon_2 = 10^{-4}$; $\bar{\psi} \leftarrow$ sufficiently large value; $\bar{\gamma} \leftarrow$ sufficiently large value; $\psi^{\text{UB}} = \bar{\psi}$, $\psi^{\text{LB}} = 0$; $\gamma^{\text{UB}} = \bar{\gamma}$, $\gamma^{\text{LB}} = 0$;
			\REPEAT
			\STATE Set $\psi = \frac{\psi^{\text{UB}} + \psi^{\text{LB}}}{2}$, $\gamma = \frac{\gamma^{\text{UB}} + \gamma^{\text{LB}}}{2}$;
			\FOR{each $k \in \mathcal{K}$}
			\STATE Calculate $a_k^*$ using \eqref{optimal_admm_a};
			\STATE Calculate $(f_k^c)^*$ using \eqref{optimal_admm_fkc};
			\ENDFOR
			
			\IF{$\sum_{k = 1}^{K} a_k^* < 1$}
			\STATE $\psi^{\text{UB}} = \psi$
			\ELSE
			\STATE $\psi^{\text{LB}} = \psi$
			\ENDIF
			\IF{$\sum_{k = 1}^{K} (f_k^c)^* < F^c$}
			\STATE $\gamma^{\text{UB}} = \gamma$
			\ELSE
			\STATE $\gamma^{\text{LB}} = \gamma$
			\ENDIF
			\UNTIL{$|\psi^{\text{UB}} - \psi^{\text{LB}}| < \varepsilon_1$ and $|\gamma^{\text{UB}} - \gamma^{\text{LB}}| < \varepsilon_2$}.
			\STATE Calculate $\tau_0^*$ using \eqref{optimal_tau_0};
			\RETURN $\{\mathbf{a}^*, (\mathbf{f}^c)^*, \tau_0^*\}$.
		\end{algorithmic}
	\end{algorithm}  \label{algorithm3}
	\subsubsection{Step 3}
	Having obtained the local and global variables $\{\mathbf{u}^{i + 1}, \mathbf{v}^{i + 1} \}$, we update the multipliers $\bm{\theta}^{i} = \{ \rho_k^i, \phi_k^i, \varphi_k^i \}$ as
	\begin{align}
		&\rho_k^{i + 1} = \rho_k^i + c \big( x_k^{i + 1} - a_k^{i + 1} \big), ~~\forall k \in \mathcal{K},  \nonumber  \\
		&\phi_k^{i + 1} = \phi_k^i + c \big(y_k^{i + 1} - (f_k^c)^{i + 1} \big), ~~\forall k \in \mathcal{K},  \nonumber  \\
		&\varphi_k^{i + 1} = \varphi_k^i + c \big(z_k^{i + 1} - \tau_0^{i + 1} \big), ~~\forall k \in \mathcal{K}.  \label{step3}
	\end{align}
	The computational complexity of Step 3 is also $O(K)$. Likewise, since \eqref{step3} can be updated in parallel for the $K$ users, the computational time of Step 3 is also constant when we perform parallel computing.
	
	We repeat the above three sequential steps until a specified stopping criterion is met. In general, the stopping criterion is specified by two thresholds: namely, an absolute tolerance $\sum_{k = 1}^{K} \big(|x_k^i - a_k^i| + |y_k^i - (f_k^c)^i| + |z_k^i - \tau_0^i| \big)$ and a relative tolerance $|\tau_0^i - \tau_0^{i - 1}| + \sum_{k = 1}^{K} \big( |a_k^i - a_k^{i - 1}| + |(f_k^c)^i - (f_k^c)^{i - 1}| \big)$ \cite{MEC_bi, ADMM}. The pseudo-code of the ADMM-based algorithm is presented in Algorithm 3. The convergence of the proposed algorithm is guaranteed as the dual problem \eqref{ADMM_dual} is convex in $\bm{\theta} = \{\bm{\rho}, \bm{\phi}, \bm{\varphi} \}$. Besides, the convergence of the algorithm is insensitive to the choice of the step size $c$ \cite{MEC_bi, ADMM2}. Without loss of generality, we simply set $c = 2$. As the complexity of each of the three steps is $O(K)$, the overall complexity of one ADMM iteration is $O(K)$. In addition, when conducting parallel computing, the computational time of the three steps are constant, thus the computational time of one ADMM iteration is constant. Therefore, the ADMM-based algorithm has excellent scalability in large-sized networks. Due to the non-convexity of (P2), a duality gap may exist and the ADMM-based algorithm may not exactly converge to the primal optimal solution of (P2). However, as we will show in the simulations, the gap between the obtained performance and the optimal one is extremely small.
	
	\begin{algorithm}[t]
		\caption{ADMM-Based Joint Service Placement and Resource Allocation Algorithm}
		\begin{algorithmic}[1]
			\STATE \textbf{initialization:} $i = 0$; $\{\bm{\rho}^i, \bm{\phi}^i, \bm{\varphi}^i\} = 0$; $b_k^i = 1$, $a_k^i = \frac{1}{K}, \forall k \in \mathcal{K}$; $h_{\text{min}} = \min\{h_k| k \in \mathcal{K}\}$, $\tau_0^i = \frac{S}{W_{\text{D}} \log_2 \big(1 + \frac{p_0 h_{\text{min}}}{N_0 W_{\text{D}}} \big)}$; $c = 2$; $\sigma_1 = 0.0005K$;
			\REPEAT
			\FOR{each user $k \in \mathcal{K}$}
			\STATE Update $\{b_k^{i + 1}, (f_k^l)^{i + 1}, x_k^{i + 1}, y_k^{i + 1}, z_k^{i + 1} \}$ by solving \eqref{step1_0} and \eqref{step1_1};
			\ENDFOR
			\STATE Update global variables $\{\mathbf{a}^{i + 1}, (\mathbf{f}^c)^{i + 1}, \tau_0^{i + 1}\}$ by solving \eqref{step2} with Algorithm 2;
			\STATE Update multipliers $\{\bm{\rho}^{i + 1}, \bm{\phi}^{i + 1}, \bm{\varphi}^{i + 1}\}$ using \eqref{step3};
			\STATE $i = i + 1$;
			\UNTIL{$ \sum\limits_{k = 1}^{K} \big(|x_k^i - a_k^i| + |y_k^i - (f_k^c)^i| + |z_k^i - \tau_0^i| \big) < 3\sigma_1$ and \\$|\tau_0^i - \tau_0^{i - 1}| + \sum\limits_{k = 1}^{K} \big( |a_k^i - a_k^{i - 1}| + |(f_k^c)^i - (f_k^c)^{i - 1}| \big) < 2 \sigma_1$};
			\RETURN $\{\mathbf{b}^i, (\mathbf{f}^l)^i, \mathbf{a}^i, (\mathbf{f}^c)^i, \tau_0^i \}$ as an approximation solution to (P2).
		\end{algorithmic}
	\end{algorithm}
	\section{Simulation Results}
	In this section, we evaluate the performance of the proposed algorithms via extensive simulations. Unless otherwise stated, we set the system uplink and downlink bandwidth as $W_{\text{U}} = W_{\text{D}} = 2$ MHz, and the noise power spectral density as $N_0 = -174$ dBm/Hz. We assume that the average channel gain $\bar{g}_k$ follows the free-space path loss model $\bar{g}_k = G \big(\frac{3 \cdot 10^8}{4 \pi f_0 d_k} \big)^{d_e}, \forall k \in \mathcal{K}$, where $G = 4.11$ denotes the antenna gain, $f_0 = 915$ MHz denotes the carrier frequency, $d_k$ denotes the distance between user $k$ and the AP, and $d_e = 3.4$ denotes the path loss exponent. The uplink channel $g_k$ follows a Rayleigh fading channel model such that $g_k = \bar{g}_k \alpha$, where $\alpha$ denotes an independent exponential random variable with unit mean. Besides, we assume that the downlink channel $h_k$ is correlated with the uplink channel $g_k$ and the correlation coefficient is set as 0.75 \cite{correlation}. Suppose that the $K$ users are located at an equal distance of 150 meters from the AP. Without loss of generality, we assume that the weighting factors are identical for all the users, i.e., $\beta_k^T = \beta$ and $\beta_k^E = 1 - \beta$, $\forall k \in \mathcal{K}$. Besides, we set equal computing energy efficiency coefficient $\kappa_k = 10^{-28}, \forall k \in \mathcal{K}$. We also assume that the computing workload is proportional to the data size, i.e., $L_k = C I_k$ \cite{workload}, where $C$ denotes the number of CPU cycles for computing one bit of task data. Unless otherwise stated, we set $K = 10$, $S = 32$ Mbits, $C = 1000$, $F^c = 20$ GHz, $p_0 = 1$ W, $F_k = 1$ GHz, $p_k = 0.1$ W, and $p_k^r = 0.01$ W \cite{circuit}, $\forall k \in \mathcal{K}$. In addition, all curves in the figures are plotted based on the average of 100 independent simulation runs, each corresponding to an independent Rayleigh fading realization.
	
	For performance comparison, we consider the following three representative benchmarks:
	\begin{enumerate}
		
		\item Optimal: the global optimal solution to (P1).
		
		\item Independent optimization: each user minimizes its own TEC independently. Specifically, the edge CPU frequency, the uplink bandwidth and downlink bandwidth are equally allocated to all the users, and each user determines whether to download the program from the AP via unicast.
		
		\item All edge computing: all the users offload their tasks to the AP for edge computing.
	\end{enumerate}
	\begin{figure}[t]
		\centering
		\includegraphics[scale=0.6]{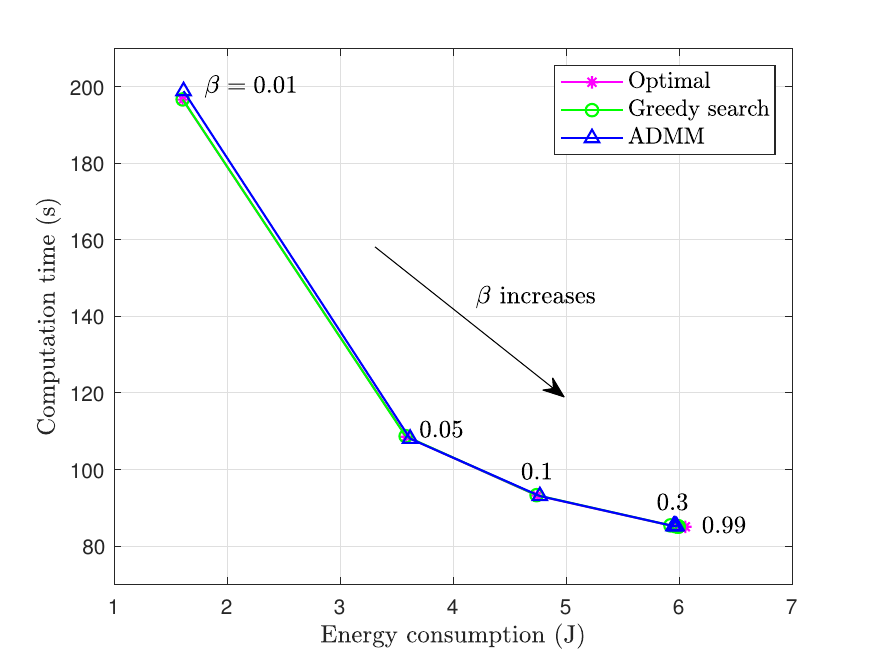}
		\caption{The optimal energy-delay tradeoff under different values of $\beta$.} \label{fig:cmp_pto_es}
	\end{figure}
	
	\subsection{TEC Performance Evaluation under Equal Task Size}
	We first consider a special case where $I_k = I$ for all the users and the $K$ users differ only by the channel gains $g_k$'s and $h_k$'s. We set $I = 8$ Mbits unless otherwise stated. In Fig. \ref{fig:cmp_pto_es}, we study the performance tradeoff between the total computation time and the total energy consumption when the weighting parameter $\beta$ varies. We can see that the performance tradeoff curves achieved by the proposed greedy search and ADMM-based algorithms are close to the optimum. Besides, we observe that as $\beta$ increases, the total computation time decreases and the total energy consumption increases. In particular, the total computation time decreases quickly with $\beta$ when $\beta$ is small and converges to a constant when $\beta \geq 0.3$. In the following simulations, we set $\beta = 0.1$ without loss of generality.
	
\begin{figure}[t]
		\centering
		\includegraphics[scale=0.6]{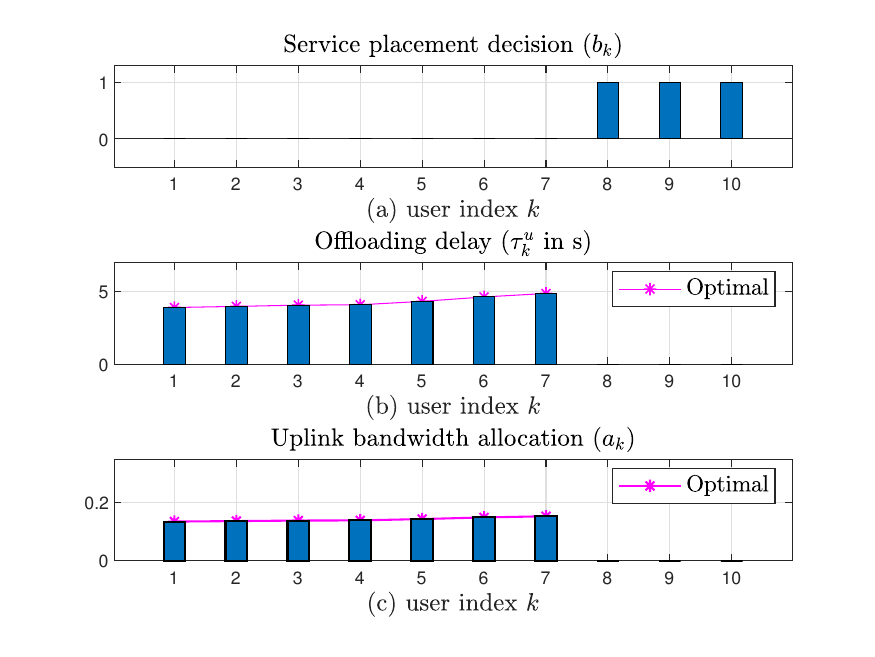}
		\caption{Optimal solution.} \label{fig:opt_p}
	\end{figure}

\begin{figure}[t]
		\centering
		\includegraphics[scale=0.6]{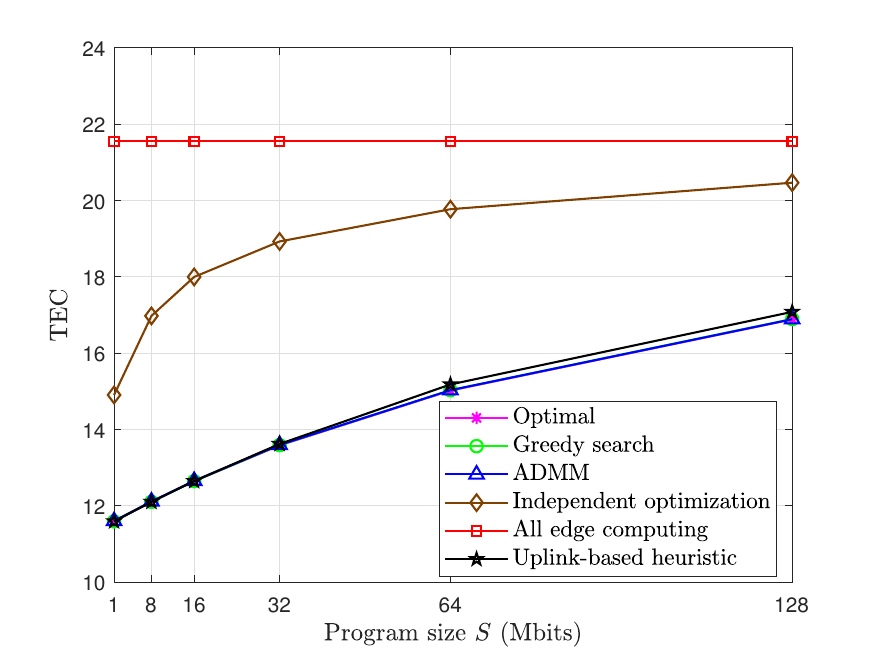}
		\caption{Total TEC versus the program size $S$.} \label{fig:cmp_S}
	\end{figure}

\begin{figure}[t]
		\centering
		\includegraphics[scale=0.6]{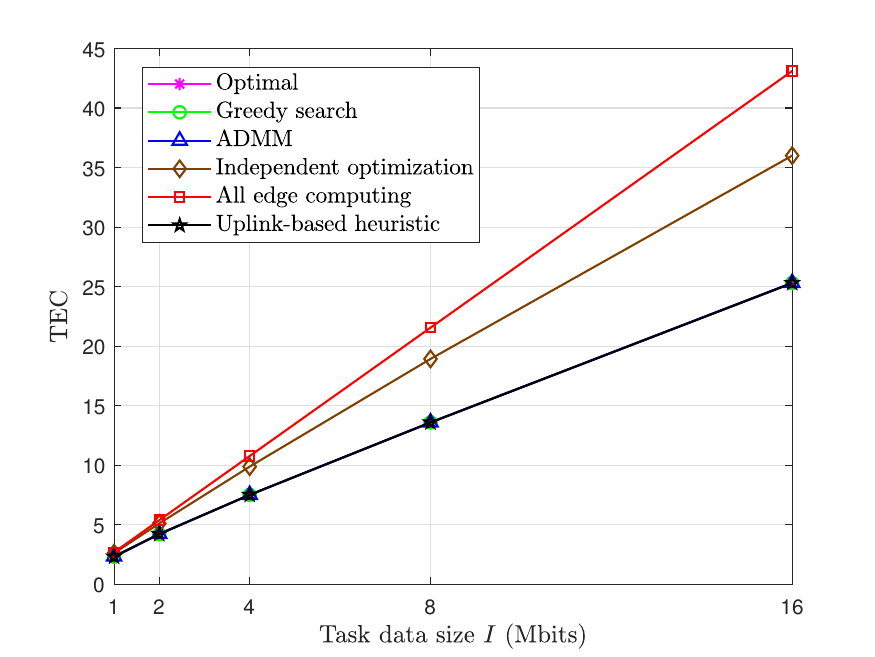}
		\caption{Total TEC versus the task data size $I$.} \label{fig:cmp_L}
	\end{figure}

\begin{figure}[t]
		\centering
		\includegraphics[scale=0.6]{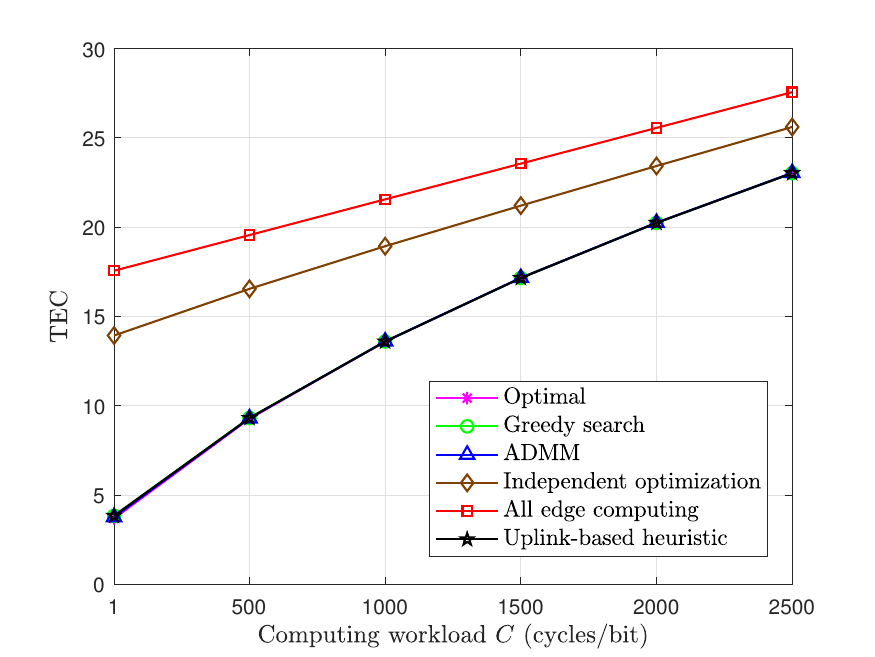}
		\caption{Total TEC versus the computing workload $C$.} \label{fig:cmp_C}
	\end{figure}

	In Fig. \ref{fig:opt_p}, we study some interesting properties of the optimal solution to Problem (P1). In particular, we sort $g_k$'s in descending order such that the uplink channel gain decreases from $g_1$ to $g_{10}$. We can see from Fig. \ref{fig:opt_p}(a) that the three users with the smallest channel gains prefer to download the service program and conduct local computing. The other users with better channels offload their tasks for edge computing. Besides, for the users in $\mathcal{K}_0$, a smaller $g_k$ leads to a longer offloading delay $\tau_k^u$, as shown in Fig. \ref{fig:opt_p}(b). The above observations verify our analysis in Section III-C. In Fig. \ref{fig:opt_p}(c), we observe that the optimal uplink bandwidth allocation $a_k$ increases when $g_k$ decreases. This indicates that more uplink bandwidth should be allocated to the users with worse channels to achieve the minimum total offloading delay. Therefore, in the following simulations in Fig. \ref{fig:cmp_S}-\ref{fig:cmp_C}, we evaluate the performance achieved by the uplink-based heuristic scheme devised in Section III-C. In this scheme, $\mathcal{K}_1$ is obtained by selecting the users in the ascending order of $g_k$.
	
	In Fig. \ref{fig:cmp_S}, we compare the TEC performance achieved by different schemes when the program size $S$ varies. Besides, we present the TEC performance comparison when the task data size $I$ varies in Fig. \ref{fig:cmp_L}. From both figures, we observe that the TEC performance achieved by the proposed greedy search and ADMM-based methods are extremely close to the optimal scheme. The three curves are on top of each other. As expected, the uplink-based heuristic scheme also achieves a close-to-optimal performance. This further confirms that users with worse uplink channels tend to conduct local computing. In addition, the proposed algorithms achieve lower total TEC than the other representative benchmarks for all values of $S$ and $I$. This demonstrates the advantage of jointly optimizing the service placement, computation offloading, and resource allocation for all the users. The total TEC increases with $S$ and $I$ for all the schemes except that the program size $S$ has no impact on the total TEC of the all-edge-computing scheme. This is because the users in the all-edge-computing scheme offload all the tasks to the edge server without downloading the service program. Besides, we observe that the proposed algorithms tend to converge to the all-edge-computing scheme when $S$ is large or $I$ is small, e.g., $I \leq 1$ Mbits. This indicates that the users tend to offload all the computation tasks to the AP when the overhead of downloading the program outweighs the gain of local computing or the offloading latency is low.

	In Fig. \ref{fig:cmp_C}, we further study the impact of the computing workload $C$ on the TEC performance when $C$ varies from $1$ to $2500$. Likewise, it can be seen that the proposed algorithms and the uplink-based heuristic scheme both achieve a close-to-optimal performance for all values of $C$. Besides, the proposed algorithms significantly outperform the other two benchmark schemes when $C$ is small, but all the schemes converge to the all-edge-computing scheme as $C$ increases. This is because when the tasks become computationally intensive, the users tend to offload the tasks to the edge server that has much more powerful computing capability.

	\subsection{TEC Performance Evaluation under Heterogeneous Task Size}
	In this subsection, we evaluate the performance of the proposed algorithms in a heterogenous case, where the users have different task data sizes $I_k$'s. In particular, the task data size of each user follows a uniform distribution with $I_k \in [1, 12]$ Mbits.
	
\begin{figure}[t]
		\centering
		\includegraphics[scale=0.6]{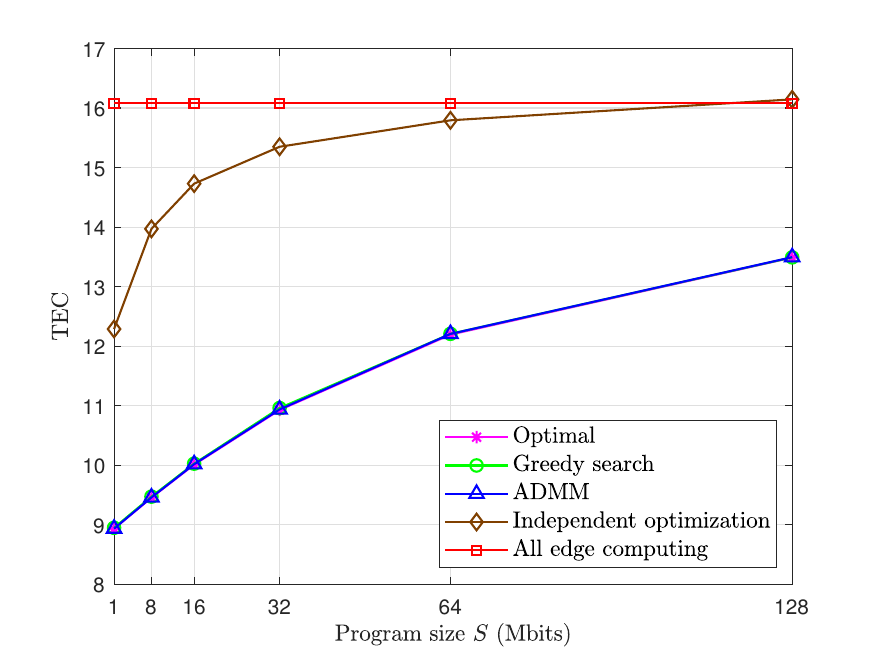}
		\caption{Total TEC versus the program size $S$.} \label{fig:cmp_S_Lh}
	\end{figure}

	In Fig. \ref{fig:cmp_S_Lh}, we plot the TEC performance achieved by different schemes when the program size $S$ increases. We observe that the performance comparison among the schemes is almost consistent with Fig. \ref{fig:cmp_S}. The proposed ADMM-based and greedy search algorithms and the optimal scheme have almost identical performance. When $S$ increases, the high downloading overhead encourages task offloading, and all the schemes converge to the all-edge-computing scheme. We can also see that the independent optimization scheme outperforms the all-edge-computing scheme when $S \leq 64$ Mbits. However, it becomes slightly worse than the all-edge-computing scheme when $S = 128$ Mbits, due to the performance loss resulted from the equal resource allocation.
	\begin{figure}[t]
		\centering
		\includegraphics[scale=0.6]{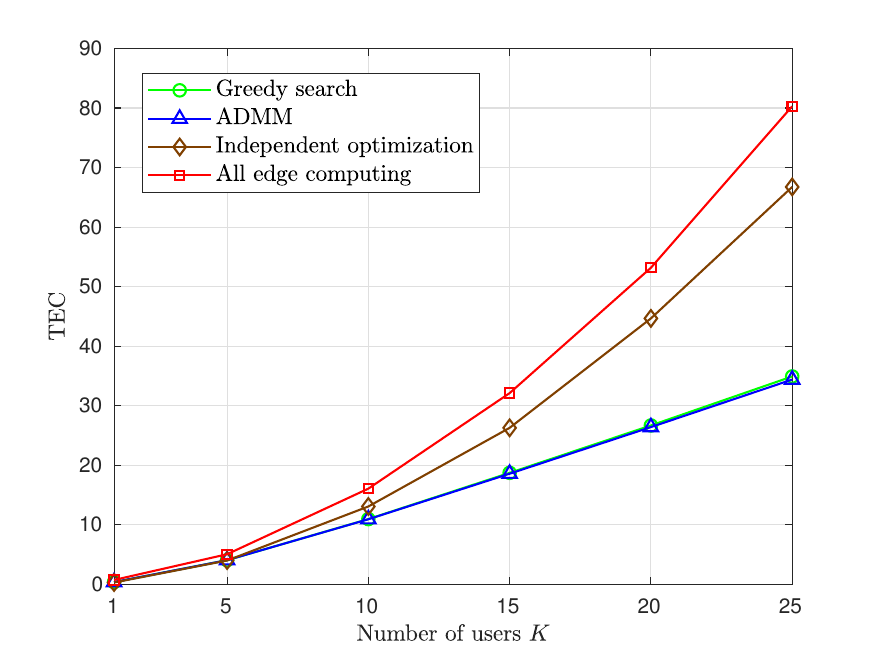}
		\caption{Total TEC versus the number of users $K$.} \label{fig:cmp_K}
	\end{figure}
	
	In Fig. \ref{fig:cmp_K}, we plot the TEC performance when the number of users $K$ varies from $1$ to $25$. Here, we do not plot the global optimal performance due to the prohibitively high computational complexity to obtain the global optimal solutions when $K$ is large. We observe that the proposed ADMM-based and greedy search algorithms have almost identical performance and outperform the benchmark schemes. The all-edge-computing scheme is the worst for all $K$ since the local computing capabilities are not utilized. When $K = 1$, the independent optimization scheme performs as well as the proposed algorithms because the single user can occupy all the resources in these three schemes. As $K$ increases, the proposed algorithms achieve increasingly lower total TEC than the two benchmarks. The reason is that the edge CPU frequency and uplink/downlink bandwidth allocated to each user becomes smaller as $K$ increases in these two benchmark schemes. Thus, users need more time to download the program or complete edge computing.
	
	\subsection{Evaluation of Computational Complexity}
\begin{figure}[t]
		\centering
		\includegraphics[scale=0.6]{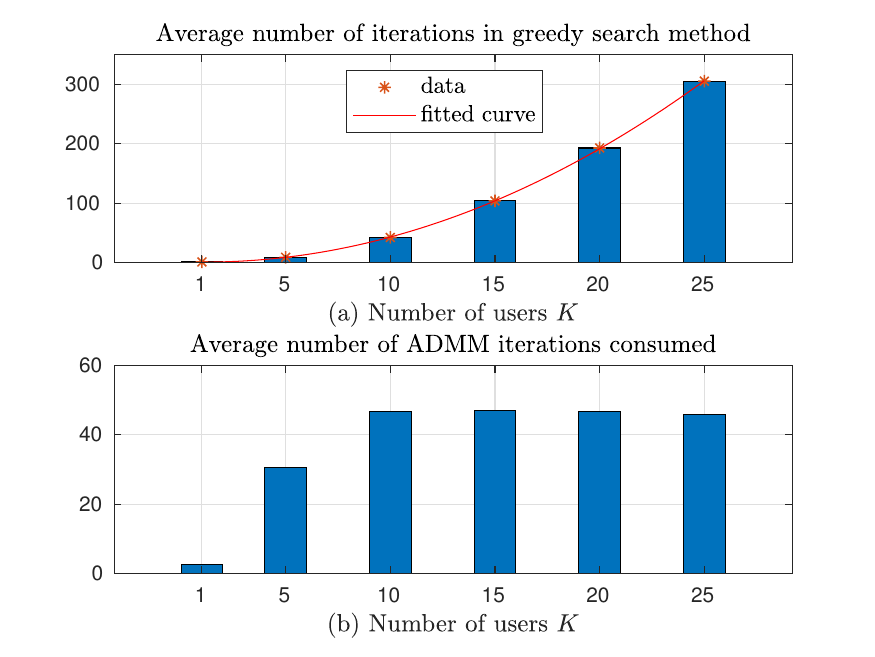}
		\caption{Computational complexity comparison between the greedy search and ADMM-based methods when $K$ varies.} \label{fig:cmp_iteration}
	\end{figure}

\begin{figure}[t]
		\centering
		\includegraphics[scale=0.6]{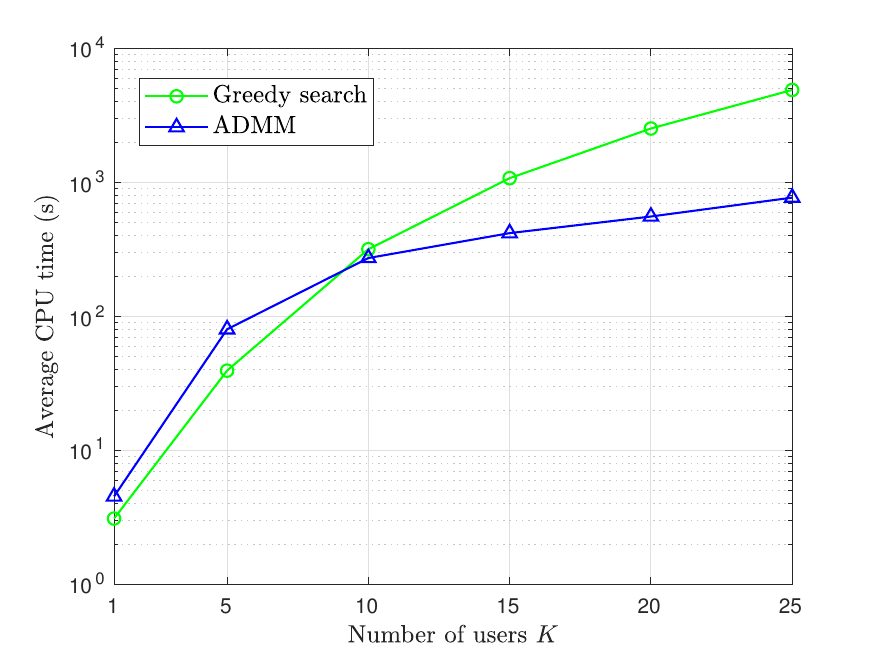}
		\caption{Average CPU time comparison between the greedy search and ADMM-based methods when $K$ varies.} \label{fig:cmp_runtime}
	\end{figure}

	In Fig. \ref{fig:cmp_iteration}, we investigate the computational complexity of the proposed greedy search and ADMM-based algorithms under the same setting of Fig. \ref{fig:cmp_K}. Here, we plot the average number of iterations consumed by the proposed greedy search method and ADMM-based method, respectively. Specifically, for each iteration in the greedy search method, the service placement decision is fixed, and Algorithm 1 is executed to solve the corresponding optimization problem in (P1). In Fig. \ref{fig:cmp_iteration}(a), we observe that the number of iterations in the greedy search method increases with $K$ at a polynomial growth rate. We also fit the number of iterations in the greedy search method to a quadratic curve, where the R-square value is 0.999962. Since each iteration corresponds to an execution of Algorithm 1, we can conclude that the number of executions of Algorithm 1 in the greedy search method scales as $O(K^2)$. Moreover, because the computational complexity of Algorithm 1 is upper bounded by $O(K^3)$, the overall computational complexity of the greedy search method is upper bounded by $O(K^5)$, which verifies our analysis in Section III-B. In Fig. \ref{fig:cmp_iteration}(b), we observe that the number of iterations consumed by the ADMM-based algorithm increases with $K$ when $K$ is small, i.e., $K \leq 10$. This is because the edge CPU frequency and uplink bandwidth allocation among the users and the service placement decisions become more flexible when $K$ increases. Thus, the ADMM-based algorithm needs more iterations to optimize the solution. When $K$ further increases, the ADMM-based algorithm takes almost a constant number of iterations, i.e., the number of ADMM iterations scales as $O(1)$. This shows that the ADMM-based algorithm can converge within a few tens of iterations \cite{ADMM}. Besides, as presented in Algorithm 3, we set the convergence threshold as $\sigma_1 = 0.0005K$, which increases with $K$ and facilitates the convergence. Furthermore, since the complexity of each iteration is $O(K)$, the overall computational complexity of the ADMM-based algorithm is $O(K)$. The above results show that the computational complexity of the ADMM-based algorithm increases much more slowly than the greedy search algorithm. In addition, we compare the average CPU time of the proposed greedy search and ADMM-based methods in Fig. \ref{fig:cmp_runtime}. In particular, we conduct parallel computing to implement the ADMM-based algorithm. Notice that the computational time of the ADMM-based algorithm still increases slightly with $K$ when $K \geq 10$ since the increase of data dimension incurs additional time for processing the data. We see that the average CPU time of the ADMM-based method is longer than the greedy search method when $K$ is small (e.g., $K \leq 8$) but increases much more slowly than the greedy search method as $K$ increases. When $K$ is large, the average CPU time of the greedy search method is much longer than the ADMM-based method. This implies that the ADMM-based algorithm is more scalable in large-sized networks.
	
	\section{Conclusions and Future Work}
	In this paper, we studied the AI service placement problem for achieving EI in a multi-user MEC system, where the edge server selectively places the AI service program at a subset of users to make use of their local computing capabilities. To minimize the computation time and energy consumption of all the users, we formulated the problem as a joint optimization of service placement, computational and radio resource allocation (on local CPU frequencies, uplink bandwidth and edge CPU frequency). We derived analytical expressions to efficiently calculate the optimal resource allocation decisions for a given service placement decision, based on which we applied search-based methods to optimize the service placement decision. We further proposed an ADMM-based algorithm to avoid high-dimensional search in large-sized networks by decomposing the original problem into parallel and tractable subproblems, one for each user. Extensive simulations showed that the proposed algorithms can achieve a close-to-optimal performance and significantly outperform various benchmark methods. By exploiting the idle local computing power, the proposed schemes significantly reduce the total computation delay and energy consumption compared with those that offload all tasks to the edge server. In particular, a larger program size leads to higher downloading delay and encourages the users to offload more tasks for edge computing. Besides, the performance advantage of the proposed algorithms becomes increasingly significant as the task data size, the weighting factor of computation time and the number of users increase, but becomes increasingly marginal as the computing workload increases.

	For a special case where the users differ only by the wireless channel gains, we observed an interesting phenomenon that the edge server tends to place the service program at the users that suffer poor channel conditions, so that the limited spectrum can be used more efficiently by the other users to offload their computation tasks. In this case, we designed a heuristic scheme based on the ascending order of uplink channel gains. Simulation results showed that the uplink-based heuristic scheme achieves a close-to-optimal performance under various system setups in such a homogeneous special case.
	
	For practical implementation, the search-based algorithms require simple calculations for the optimal resource allocation decisions with the derived analytical expressions. However, the computational complexity of the ADMM-based algorithm increases linearly with the network size, which has a much smaller growth rate than the search-based algorithms. Besides, with parallel computing, the computational time of the ADMM-based algorithm increases much more slowly with the network size. Therefore, the search-based algorithms are preferred when the network size is small and the proposed ADMM-based algorithm is much more scalable in large-sized networks. The proposed optimization framework in this paper can be leveraged to achieve communication-efficient edge-device inference, which is an important part of the vision on 6G communications, i.e., to support ubiquitous AI services with limited communication, computation, hardware, and energy resources \cite{6G}.
	
	Finally, we conclude the paper with some interesting future directions. First, it is interesting to consider service placement and resource allocation in a general multi-AP MEC system, where multiple APs collaboratively serve users to provide better and more reliable system computing performance. In this case, different communication protocols, e.g., orthogonal frequency-division multiple access (OFDMA), nonorthogonal multiple access (NOMA), and time division multiple access (TDMA), can be utilized to address the new challenges of AP-user association and interference management. Second, it is also promising to extend the single-service setup to a multi-service one. The system resources can be shared by all services or split into separate parts through virtualization techniques, one for each service. Meanwhile, the edge servers can achieve load balance by adaptively allocating the computing resources. Moreover, we assumed in this paper that the channel conditions and computation requirements are known in advance and studied an offline optimization problem. In practice, users are dynamic and may request the AI service at different time and with different frequencies. Hence, an online design is needed to apply to time-varying channel conditions and computation requirements.
	
	\bibliography{references}
\end{document}